\documentclass[12pt]{article}
\usepackage{amsfonts}
\usepackage{amssymb}
\usepackage{latexsym}
\usepackage{graphicx}
\usepackage[english]{babel}
\begin{document}
\input epsf


\begin{flushright}
hep-th/0502050
\end{flushright}
\vspace{20mm}
 \begin{center}
{\LARGE  The fuzzball proposal for black holes:\\ an elementary review\footnote{Lecture given at the RTN workshop `The quantum structure of space-time and the geometric nature of fundamental interactions', in Crete, Greece  (September 2004). }}
\\
\vspace{18mm}
{\bf   Samir D. Mathur
\\}
\vspace{8mm}
Department of Physics,\\ The Ohio State University,\\ Columbus,
OH 43210, USA\\ 
\vspace{4mm}
 {mathur@mps.ohio-state.edu}\\
\vspace{4mm}
\end{center}
\vspace{10mm}
\thispagestyle{empty}
\begin{abstract}

We give an elementary review of black holes in string theory. We discuss BPS holes,
the microscopic computation of entropy and the `fuzzball' picture of the black hole interior
suggested by microstates of the 2-charge system.

\end{abstract}
\newpage
\renewcommand{\theequation}{\arabic{section}.\arabic{equation}}

\def\p{\partial}
\def\be{\begin{equation}}
\def\ee{\end{equation}}
\def\bea{\begin{eqnarray}}
\def\eea{\end{eqnarray}}
\def\r{\rightarrow}
\def\tildr{\tilde}
\def\n{\nonumber}
\def\nn{\nonumber \\}

\section{Introduction}

The quantum theory of black holes presents many paradoxes. It is vital to ask how these paradoxes are to be resolved,
for the answers will likely lead to deep changes in our understanding of quantum gravity, spacetime and matter.

Bekenstein \cite{bek} argued that black holes should be attributed an entropy
\be
S_{Bek}={A\over 4G}
\label{sevent}
\ee
where $A$ is the area of the horizon and $G$ is the Newton constant of gravitation. (We have chosen units to set $c=\hbar=1$.) This entropy must be attributed to the hole if we are to prevent a violation of the second law of thermodynamics. We can throw  a box of gas with  entropy $\Delta S$ into a black hole, and see it vanish into the central singularity. This would seem to decrease the entropy of the Universe, but we note that the area of the horizon increases as a result of the energy added by the box. It turns out that if we assign (\ref{sevent}) as the entropy of the hole then the total entropy is nondecreasing
\be
{dS_{Bek}\over dt}+{dS_{matter}\over dt}\ge 0
\ee
This would seem to be a good resolution of the entropy problem, but it leads to another problem. The principles of statistical mechanics tell us that there must be
\be
{\cal N}=e^{S_{Bek}}
\ee
microstates of the black hole. But traditional attempts to find these  microstates did 
not succeed; rather it appeared that the black hole geometry was uniquely determined by the conserved charges of the hole.
In colloquial terms, `Black holes have no hair'. 

One may think that the differences between the $e^{S_{Bek}}$ microstates of the hole are to be found by looking at a planck sized neighborhood of the singularity, and these differences are thus not visible in the classical description. After all, the matter that made up the hole disappeared into the singularity at $r=0$. But this picture of the hole leads to a much more serious problem, the `information paradox'. Hawking showed that vacuum modes near the horizon evolve into particle pairs; one member of the pair falls into the hole and reduces its mass, while the other escapes to infinity as `Hawking radiation' \cite{hawking}. 
If the information about the microstate resides at $r=0$ then the outgoing radiation is insensitive to the details of the microstate, and when the hole has evaporated away  we cannot recover the information contained in the matter which went in to make the hole. This is a violation of the unitarity of quantum mechanics, and thus a severe contradiction with the way we understand evolution equations in physics. 

The information paradox has resisted attempts at resolution for some 30 years. The robustness of the paradox stems from the fact that it uses very few assumptions about the physics relevant to the hole. One assumes that quantum gravity effects are confined to a small length scale like the planck length or string length, and then notes that the curvature scales at the horizon are much larger than this length for large black holes. Thus it would appear that the precise theory of quantum gravity is irrelevant to the process of Hawking radiation and thus for the resolution of the paradox. 

String theory is a consistent theory of quantum gravity; further, it is a theory with no free parameters. We should therefore ask 
how this theory deals with black holes. The past decade has shown dramatic progress in our understanding of black holes in string theory. We have understood how to count microstates of black holes. Recent computations suggest that the resolution of the information paradox lies in the fact that quantum gravity effects do not stay confined to microscopic distances, and the black hole interior is quite different from the naive picture suggested by classical gravity. This review gives an elementary introduction to these ideas and conjectures. 

     \begin{figure}[htbp]
   \begin{center}
   \includegraphics[width=4in]{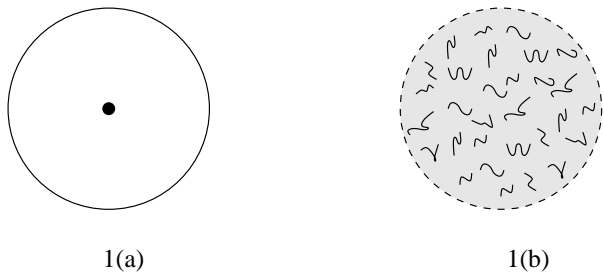}
   \caption{(a) The conventional picture of a black hole \quad (b) the proposed picture -- state information is distributed throughout the `fuzzball'. }
   \label{fig1}
   \end{center}
   \end{figure}

\section{Making black holes in string theory}

Susskind et. al \cite{susskind} proposed an interesting approach to studying black holes in string theory. Consider a highly excited state of a string. Thus the state has a  mass $M\gg \alpha'^{-1/2}$. Assume that the string coupling is small ($g\ll1$) so the string is essentially free. The left and right oscillator levels are $ N_L, N_R\sim \sqrt{\alpha'} M\gg 1$, so there is a large degeneracy ${\cal N}$ of states with this mass. This count of microscopic states gives an entropy $S_{micro}= \ln[{\cal N}]\sim \sqrt{\alpha'} M$. (The exact proportionality constant depends on how many directions are compactified; such compact directions provide winding modes that also contribute to the entropy.) 

Now imagine increasing the string coupling $g$; this brings in gravity since the Newton gravitational constant is $G\sim  g^2$. If $M$ was sufficiently large then we expect to get a {\it black hole} of mass $M$. We can compute the Bekenstein entropy of this hole $S_{Bek}=A/4G$. For a Schwarzschild hole in 3+1 noncompact dimensions we get $S_{Bek}\sim M^2$. More generally, if we had $D$ noncompact spacetime directions we get $S_{Bek}\sim M^{D-2\over D-3}$. 

Thus $S_{micro}$ and $S_{Bek}$ grow as different powers of $M$, and it might seem that we have learnt nothing. But it is nontrivial that   $S_{micro}$ grows as a power of $M$ at all, since this implies that even at $g=0$ the microscopic state count grows exponentially with energy.  This exponential growth can be traced back to the fact that we had an {\it extended} object (the string) into which we were putting the energy $M$ -- if we had a theory containing point particles only then at $g=0$ 
we expect that these particles would disperse rather than form a highly degenerate bound state.

\subsection{BPS states}

The discrepancy between $S_{micro}$ and $S_{Bek}$ must be due to the fact that energy levels shift as we change $g$, which makes it incorrect to directly compare degeneracies of states at different $g$. We could avoid this problem if we look at BPS states of string theory. These states have  charge as well as mass, with $Q=M$ in suitable units. The mass of a BPS state is given by the charges it carries and the values of the moduli in the theory. So all BPS states with a mass $M$ at a given value of $g$ move together as we change $g$ and we can compare degeneracies at different values of $g$. We therefore expect that
\be
S_{micro}=S_{Bek}
\label{one}
\ee
for BPS states. In fact if we were to {\it not} get such an agreement then we would be seriously worried about string theory as a theory of quantum gravity. The entropy $S_{Bek}$ was obtained by thermodynamic arguments, and it is the task of the full theory to reproduce this entropy through a microscopic count of states. Conversely, if we {\it do} get the agreement (\ref{one}) then string theory would have passed a very nontrivial test in that it can claim to have the right degrees of freedom required from a theory of quantum gravity. 

\subsection{The 1-charge solution}

Let us start by looking at the simplest BPS states in the theory. Let us take type IIA string theory for concreteness; we can also understand this as the dimensional reduction of 11-D M theory. Let us compactify a circle $S^1$  in the IIA theory; we label this circle by a coordinate $y$ with
\be
0\le y<2\pi R
\ee
We can wrap an elementary string (NS1 brane) around this circle. With no oscillators excited on the string, this is a BPS state. We can also consider a string wrapped $n_1$ times around the circle. With $n_1$ large this will give us a BPS state with large mass. The supergravity solution produced by such a string is
\bea
ds^2_{string}&=&H_1^{-1}[-dt^2+dy^2]+\sum_{i=1}^8 dx_idx_i\\
e^{2\phi}&=&H_1^{-1}\\
H_1&=&1+{Q_1\over r^6}
\eea
Here $ds^2_{string}$ is the 10-D string metric and $x_i$ are the 8 spatial directions transverse to the string. At $r\rightarrow 0$ the dilaton $\phi$ goes to $-\infty$  and the length of the $y$ circle is seen to go to zero. The geometry does not have a horizon at any nonzero $r$, and if we say that the horizon occurs at $r=0$ then we find that the area of this horizon (measured in the Einstein metric) is zero. Thus we get $S_{Bek}=0$.

This vanishing of $S_{Bek}$ is actually consistent with the microscopic  count. The NS1 brane is in an oscillator ground state, so its only degeneracy comes from the zero modes of the string, which give 128 bosonic and 128 fermionic states. Thus we have $S_{micro}=\ln[256]$ which does not grow with   $n_1$. Thus in the macroscopic limit $n_1\rightarrow \infty$ we would write $S_{micro}=0$ to leading order, which agrees with $S_{Bek}$.

Let us go back and see why we failed to make a  black hole with nonzero area. Consider the NS1 brane as an M2 brane of M theory; this M2 brane wraps the directions $x_{11},y$. A brane has tension along its worldvolume directions, so it squeezes the cycles on which it is wrapped. Thus the length of the $x_{11}$ circle goes to zero at the brane location $r=0$, which shows up as $\phi\rightarrow -\infty$ in the IIA description. Similarly, we get a vanishing of the length of the $y$ circle in the M theory description. On the other hand if we had some directions that are compact and transverse to a brane then they would tend to expand; this happens because the flux radiated by the brane has  energy  and this energy  is lower if the flux is spread over a larger volume. 

In computing the area of the horizon we can take two equivalent approaches: 

(a) We can just look at the D noncompact directions, and find the Einstein metric (after dimensional reduction) for these noncompact directions. We compute the area $A_D$ in this metric and use the Newton constant $G_D$ for $D$ dimensions to get $S_{Bek}=A_D/4G_D$. .

(b) We can compute the area of the horizon in the full 11-D metric of M theory, and use the Newton constant for 11-D to get
$S_{Bek}=A_{11}/4 G_{11}$. In the IIA description we can compute the area of the horizon in the 10-D Einstein metric and write $S_{Bek}=A^E_{10}/4 G_{10}$. 

It is easy to check that the two computations give the same result. Let us follow (b). Then we can see that the vanishing of the $x_{11}$ and $y$ circles in the above 1-charge solution will make the 11-D horizon area vanish, and give $S_{Bek}=0$.

\subsection{Two charges}

To avoid the shrinking of the direction $x_{11}$ we can take  M5 branes and place them transverse to the direction $x_{11}$; this gives  NS5 branes in the IIA theory. To wrap the  five spatial worldvolume directions of  the NS5 branes we need more compact directions, so let us compactify a $T^4$ in addition to the $S^1$  and wrap the NS5 branes on this $T^4\times S^1$. We still have the NS1 branes along $y$, but note that with the additional compactifications the power of $r$ occurring in $H_1$ changes. We get
\bea
ds^2_{string}&=&H_1^{-1}[-dt^2+dy^2]+H_5\sum_{i=1}^4 dx_idx_i+\sum_{a=1}^4 dz_adz_a\nn
e^{2\phi}&=&{H_5\over H_1}\nn
H_1&=&1+{Q_1\over r^2}, ~~~~~\qquad H_5=1+{Q_5\over r^2}
\eea
 The $T^4$ is parametrized by $z_a, a=1\dots 4$. $Q_5$ is proportional to $n_5$, the number of NS5 branes. Note that the dilaton stabilizes to a  constant as $r\rightarrow 0$; this reflects the stabilization of the $x_{11}$ circle. Note that the $T^4$ also has a finite volume at $r=0$ since the NS5 branes cause it to shrink (their worldvolume is along the $T^4$) while the NS1 branes cause it to expand (they are transverse to the $T^4$). But the horizon area in the Einstein metric  is still zero; this can be seen in the M theory description from the fact that the NS1(M2) and the NS5 (M5)  both wrap the $y$ circle and cause it to shrink to zero at $r=0$.

\subsection{Three charges}
 
 To stabilize the $y$ circle we add {\it momentum} charge P along the $y$ circle. If we have $n_p$ units of momentum along $y$ then the energy of these modes is $n_p/R$, so their energy is {\it lower} for larger $R$. This contribution will therefore counterbalance the effect of the NS1, NS5 branes for which the energies were linearly proportional to $R$. We get
 \bea
ds^2_{string}&=&H_1^{-1}[-dt^2+dy^2+K(dt+dy)^2]+H_5\sum_{i=1}^4 dx_idx_i+\sum_{a=1}^4 dz_adz_a\n \\ 
e^{2\phi}&=&{H_5\over H_1}\n \\
H_1&=&1+{Q_1\over r^2}, ~~~~~~~~~~\qquad H_5=1+{Q_5\over r^2}, ~~~~~~~~~~~\qquad K={Q_p\over r^2}
\label{tenp}
\eea
This metric has a horizon at $r=0$. We will compute the area of this horizon in the 10-D string metric, and then convert it to the area in the Einstein metric. 

 Let us write the  metric in the noncompact directions  in polar coordinates and examine it near $r=0$
 \be
 H_5 \sum dx_idx_i= H_5(dr^2+r^2d\Omega_3^2)\approx Q_5[{dr^2\over r^2}+d\Omega_3^2]
 \ee
 Thus the area of the transverse $S^3$ becomes a constant at $r\r 0$ 
 \be
 A_{S^3}^{string} = (2\pi^2)Q_5^{3\over 2}
 \ee
 The length of the $y$ circle at  $r\r 0$ is
 \be
 L_y^{string}=(2\pi R) ({K\over H_1})^{1\over 2}= 2\pi R {Q_p^{1\over 2}\over Q_1^{1\over 2}}
 \ee
 Let the coordinate volume spanned by the $T^4$ coordinates $z_a$ be $(2\pi)^4 V$. The volume of $T^4$ at $r\rightarrow 0$ is
 \be
 V_{T^4}^{string}=(2\pi)^4 V 
 \ee
 Thus the area of the horizon at $r=0$ is
 \be
 A^{string}= A_{S^3}^{string} L_y^{string}V_{T^4}^{string}=(2\pi^2)(2\pi R)((2\pi)^4 V)Q_1^{-{1\over 2}}Q_5^{3\over 2}Q_p^{1\over 2}
 \ee
The 10-D Einstein metric $g^E_{ab}$ is related to the string metric $g^S_{ab}$ by
\be
g^E_{ab}=e^{-{\phi\over 2}}g^S_{ab}={H_1^{1\over 4}\over H_5^{1\over 4}}g^S_{ab}
\ee
At $r\rightarrow 0$ we have $e^{2\phi}={Q_5\over Q_1}$, which gives for the area of the horizon in the Einstein metric
\be
A^E=({g^E_{ab}\over g^S_{ab}})^4A^{string}={Q_1\over Q_5}A^{string}=(2\pi^2)(2\pi R)((2\pi)^4 V)(Q_1Q_5Q_p)^{1\over 2}
\ee
The Newton constant of the 5-D noncompact space $G_5$ is related to the 10-D Newton constant $G_{10}$ by
\be
G_5={G_{10}\over (2\pi R) ((2\pi)^4 V)}
\ee
We can thus write the Bekenstein entropy as
\be
S_{Bek}={A^E\over 4G_{10}}={(2\pi^2)(2\pi R)((2\pi)^4 V)(Q_1Q_5Q_p)^{1\over 2}\over 4 G_{10}}={(2\pi^2)(Q_1Q_5Q_p)^{1\over 2}\over 4G_5}
\label{two}
\ee
We next express the $Q_i$ in terms of the integer charges
\bea
Q_1&=& { g^2 \alpha'^3\over V} n_1\nn
Q_5&=&\alpha' n_5\nn
Q_p&=&{g^2\alpha'^4\over VR^2}n_p
\label{sixt}
\eea
We have
\be
G_{10}=8\pi^6 g^2\alpha'^4
\ee
Substituting in (\ref{two}) we find
\be
S_{Bek}=2\pi(n_1n_5n_p)^{1\over 2}
\label{threeqp}
\ee
Note that  the moduli $g, V, R$ have all cancelled out. This fact is crucial to the possibility of reproducing this entropy by some microscopic calculation. In the microscopic description we will have a bound state of the charges $n_1, n_5, n_p$ and we will be counting the degeneracy of this bound state. But since we are looking at BPS states this degeneracy will not depend on the moduli. 

\subsection{Dualities}

We have used three charges above: NS1 branes wrapped on $S^1$, NS5 branes wrapped on $T^4\times S^1$, and momentum P along $S^1$. If we do a T-duality in one of the directions of $T^4$ we do not change any of the charges, but reach type IIB string theory. We can now do an S-duality which gives
\be
NS1\, NS5\, P~\stackrel{\textstyle S}{\rightarrow} ~ D1\, D5\,P
\ee
Historically the D1-D5-P system was studied first, and so for many purposes we will work with that system. Note that dualities can also be used to permute the three charges among themselves in all possible ways. Four T-dualities along the four $T^4$ directions will interchange the D1 with the D5, leaving P invariant. Another set of dualities can map D1-D5-P to P-D1-D5, which after S-duality gives P-NS1-NS5. Since we will make use of this map later, we give it explicitly here (the direction $y$ is called $x^5$ and the $T^4$ directions are called $x^6\dots x^9$)
\bea
&D1\, D5\,P\, (IIB)&~ \stackrel{\textstyle S}{\rightarrow}~NS1\, NS5\, P\,(IIB) \nonumber \\
&&~\stackrel{\textstyle T_5}{\rightarrow}~ P\, NS5 \, NS1\, (IIA)\nonumber \\
&&~\stackrel{\textstyle T_6}{\rightarrow}~ P\, NS5 \, NS1\,(IIB)\nonumber \\
&&~\stackrel{\textstyle S}{\rightarrow}~ P\, D5\, D1\,(IIB)\nonumber \\
&&\stackrel{\textstyle T_{6789}}{\rightarrow}~ P\, D1\, D5\,(IIB)\nonumber \\
&&~\stackrel{\textstyle S}{\rightarrow}~ P\, NS1 \, NS5 \,(IIB)
\label{twop}
\eea
If we keep only the first two charges in the above sequence then we see that the D1-D5 bound state is dual to the P-NS1 state. This duality will help us understand the geometric structure of the D1-D5 system, since P-NS1 is just an elementary string carrying vibrations.

\section{The microscopic count of states}

We have already seen that for the one charge case (where we had just the string NS1 wrapped on a circle $n_1$ times) we get
$S_{micro}=\ln[256]$. This  entropy does not grow with the winding number $n_1$ of the string, so from a macroscopic perspective we  get $S_{micro}\approx 0$. 

Let us now consider two charges, which we take to be NS1 and P; by the above described dualities this is equivalent to taking any two  charges from the set NS1-NS5-P or from the set D1-D5-P. The winding number of the NS1 is $n_1$ and the number of units of momentum is $n_p$. It is important that we consider the {\it bound} state of all these charges. If we break up the charges into two separate bound states then we would be describing {\it two} black holes rather than the single black hole that we wish to consider. 

\subsection{Bound states and unbound states}

Since this issue is important, let us discuss this in some more detail. Suppose we take two D-p branes, and wrap them on a p-torus  (which has volume $V_p$). Thus the two D-p branes are parallel to each other, but we separate them by a distance $L$ in the noncompact directions. Is this a BPS configuration and one that we must consider when computing degeneracies of states in the black hole context?

Each D-p brane has a mass $M=T_p V_p$, where $T_p$ is the brane tension. If we localize the centers of the  branes too sharply, then there will be a high value of the quantum `localization' energy. Let $x$  denote the relative separation  of the branes. Since we wish to place the branes a distance $L$ apart, we would need to define the positions upto some accuracy $\Delta x=\mu L$, with $\mu \ll1$. This gives for the conjugate momentum $\Delta p
 \sim 1/(\mu L)$, and thus  a localization energy 
 \be
 \Delta E\sim {(\Delta p)^2\over 2M}\sim {1\over M \mu^2 L^2}
 \ee 
 
 Note that for a D-brane, $M\sim g^{-1}(\alpha')^{-{p+1\over 2}}V_p$ so 
 \be
 \Delta E\sim {g(\alpha')^{p+1\over 2}\over V_p\mu^2 L^2}
 \ee 
 In the limit $g\rightarrow 0$ we might ignore this energy, but for any finite $g$ this localization energy $\Delta E$ adds on to the BPS mass carried by the  branes and makes the overall  energy greater than charge. Thus we do not get a BPS state. It is often said that D-p branes placed parallel to each other are in a BPS configuration, but this can be true only if the branes are either infinite in extent ($V_p=\infty$) or if $g=0$. For the application to black holes, neither of these is the case.
 
 How then do we get a BPS state from parallel D-p branes? If we place the branes very close to each other then we get the above mentioned localization energy, but the quantum fluctuations also give $\Delta p\ne 0$. Branes that are moving attract each other. By a delicate cancellation of this latter effect against the energy required for localization we can get a D-p brane bound state, and this state is BPS. The count of such states is known -- for example if we have any number $N$ of D-p branes (wrapped on a p-torus) then they form a bound state with degeneracy 256; this degeneracy is caused by fermion zero modes on the branes. We do not get additional BPS states by putting the individual branes at arbitrary spatial locations. Even though the localization energy can be small, we need to be careful to count only the true BPS states since we will be doing dualities which take us to various domains of parameter space, and the energy corrections for non-BPS states need not be ignorable in all of these domains.
 
\subsection{Entropy of the NS1-P bound state}
 
 Let us return to the 2-charge NS1-P state. For this set of two charges it is easy to identify what states are bound. First we must join all the windings of the NS1 together to make a single string; this `long string' loops $n_1$ times around $S^1$ before joining back to its starting point. The momentum P must also be bound to this long string. If the momentum was {\it not} bound to the NS1 then it would manifest itself as massless quanta of the IIA theory (graviton, gauge fields etc) rotating around the $S^1$. When the momentum P is {\it bound} to the NS1 then it takes the form of traveling waves on the NS1. Thus the bound state of NS1-P is just a single `multiwound' string wrapped on $S^1$ with waves traveling in one direction along the $S^1$.  

But there are several ways to carry the same momentum P on the NS1. We can partition the momentum among different harmonics of vibration, and this gives rise to a large number of states for a given choice of $n_1, n_p$. We will count the degeneracy in three closely related ways; each way will help us understand some facet of the degeneracy later on.

\medskip

(a) One way to count the states is to note that we just have the elementary string carrying winding number $n_1$, momentum $n_p$, left moving excitations at some level $N_L$, but no right moving oscillations $N_R$. (We set $N_R=0$ so that the string state maintains the supersymmetries coming from the right moving sector; this makes the state BPS \cite{sen}.) The mass of a string  state is given by
\be
m^2=(2\pi R n_1 T -{n_p\over R})^2+8\pi T N_L=(2\pi R n_1 T +{n_p\over R})^2+8\pi T N_R
\ee
where $R$ is the radius of the $S^1$ and $T$ is the tension $1/(2\pi\alpha')$ of the elementary string. With $N_R=0$ we get
\be
N_L=n_1n_p
\ee
and
\be
m=2\pi R n_1 T +{n_p\over R}
\ee
so we have a threshold bound state of the NS1 and P charges (i.e. there is no binding energy). Since we will take $n_1, n_p$ to be macroscopically large, we have $N_L\gg1$. This oscillator level is partitioned between 8 bosonic oscillators (which together give a central charge $c=8$) and 8 fermionic oscillators (which together have central charge $c=8/2=4$). Thus the total central charge is
\be
c=8+4=12
\ee
The number of states at oscillator level $N_L$ is given by
\be
{\cal N}\sim e^{2\pi\sqrt{{c\over 6}N_L}}=e^{2\sqrt{2}\pi\sqrt{n_1n_p}}
\ee
The entropy is thus
\be
S_{micro}=\ln{\cal N}=2\sqrt{2}\pi\sqrt{n_1n_p}
\label{three}
\ee

\medskip

(b) Since we are looking at BPS states, we do not change the count of states by taking $R$ to be vary large. In this
limit we have small transverse vibrations of the NS1. We can take the DBI action for the NS1, choose the static gauge, and obtain an action for the vibrations that is just quadratic in the amplitude of vibrations. The vibrations travel at the speed of light along the direction $y$. Different Fourier modes separate and  
each Fourier mode is described by a harmonic oscillator. The total length of the NS1 is
\be
L_T=2\pi R n_1
\ee
Each excitation of the Fourier mode $k$ carries energy and momentum
\be
e_k=p_k={2\pi k\over L_T}
\ee
The total momentum on the string can be written as
\be
P={n_p\over R}={2\pi n_1n_p\over L_T}
\ee
First focus on only one of the transverse directions of vibration. If there are $m_i$ units of the Fourier harmonic $k_i$ then we need to have
\be
\sum_i m_ik_i=n_1n_p
\ee
Thus the degeneracy is given by counting  {\it partitions} of the integer $n_1n_p$.  The number of such partitions is known to be
$\sim Exp(2\pi\sqrt{n_1n_p\over 6})$. We must however take into account the fact that the momentum will be partitioned among 8 bosonic vibrations and 8 fermionic ones; the latter turn out to be equivalent to 4 bosons. Thus there are ${n_1n_p\over 12}$ units of momentum for each bosonic mode, and we must finally multiply the degeneracy in each mode. This gives
\be
{\cal N}=[Exp(2\pi\sqrt{n_1n_p\over 72})]^{12}=Exp(2\pi\sqrt{2}\sqrt{n_1n_p})
\ee
which again gives the entropy (\ref{three}).

\medskip

(c) We can look at the vibrations of (b) above as a 1-dimensional {\it gas} of massless quanta traveling on the NS1 string. 
The gas lives in a `box' of length $L_T=2\pi R n_1$. All quanta in the gas travel in the same direction, so the gas has a total energy and momentum
\be
E=P={n_p\over R}={2\pi n_1n_p\over L_T}
\ee
Further there are 8 bosonic degrees of freedom and 8 fermionic degrees of freedom. We  can write a partition function $Z$  for the bosonic and fermionic modes
\be
Z=\sum_{states} e^{-\beta E_{state}}
\ee
For a bosonic mode of harmonic $k$ each quantum has energy $e_k={2\pi k/L_T}$, so its contribution to the partition function is
\be
Z_k^B\rightarrow \sum _{m_k=0}^\infty e^{-\beta m_ke_k}={1\over 1-e^{-\beta e_k}}
\ee
Similarly a fermionic mode in the harmonic $k$ contributes
\be
Z_k^F\rightarrow \sum _{m_k=0}^1 e^{-\beta m_ke_k}={ 1+e^{-\beta e_k}}
\ee
We consider the log of $Z$, so that we add the logs of the individual contributions above.  We then approximate the sum over $k$ by an integral ($\sum_k\rightarrow \int dk = {L_T\over 2\pi}\int d e_k$)
getting for bosonic modes
\be
\log Z^B\rightarrow -{L_T\over 2\pi}\int _0^\infty de_k \ln [ 1-e^{-\beta e_k}]={L_T\over 2\pi \beta }{\pi^2\over 6}
\ee
and for fermionic modes
\be
\log Z^F\rightarrow {L_T\over 2\pi}\int _0^\infty de_k \ln [ 1+e^{-\beta e_k}]={L_T\over 2\pi \beta }{\pi^2\over 12}
\ee
If we have $f_B$ bosonic degrees of freedom and $f_F$ fermionic degrees of freedom we get
\be
\log Z=(f_B+{1\over 2} f_F){\pi L_T\over 12 \beta}\equiv c({\pi L_T\over 12 \beta})
\label{five}
\ee
We can see explicitly in this computation that a fermionic degree of freedom counts as half a bosonic degree of freedom.
From the 8 transverse bosonic vibrations and 8 fermionic vibrations we get $c=12$.

We determine $\beta$ by
\be
E=-\partial_\beta (\ln Z)={c\pi L_T\over 12 \beta^2}
\label{four}
\ee
which gives for the temperature
\be
T=\beta^{-1} =[{12 E\over \pi L_T c}]^{1\over 2}
\ee
The entropy is
\be
S=\ln Z+\beta E={c\pi L_T\over 6\beta}=[{c\pi L_T E\over 3}]^{1\over 2}
\label{sixqq}
\ee
Substituting the values of $c,E$ we again find
\be
S=2\sqrt{2}\pi\sqrt{n_1n_p}
\ee
From the above computation we can however extract a few other details.  The average energy of a quantum
will be 
\be
e\sim T\sim {\sqrt{n_1n_p}\over L_T}
\ee
so that the generic quantum is in a harmonic 
\be
k\sim \sqrt{n_1n_p}
\label{thir}
\ee
on the multiwound NS1 string. Given that the total energy is (\ref{four}) we find that the number of such quanta is
\be
m\sim \sqrt{n_1n_p}
\ee
The occupation number of an energy level $e_k$ is
\be
<m_k>={1\over 1-e^{-\beta e_k}}
\ee
so for the generic quantum with $e_k\sim \beta^{-1}$ we have
\be
<m_k>\sim 1
\label{ttseven}
\ee

\bigskip

To summarize, there are a large number of ways to partition the energy into  different harmonics. One extreme possibility is to put all the energy into the lowest harmonic $k=1$; then the occupation number of this harmonic will be
\be
m=n_1n_p
\ee
At the other extreme we can put all the energy into a single quantum in the harmonic $n_1n_p$; i.e.
\be
k=n_1n_p, ~~~m_k=1
\ee
But the {\it generic} state which contributes to the entropy has its typical excitations in harmonics with  $k\sim \sqrt{n_1n_p}$. There are
$\sim\sqrt{n_1n_p}$ such modes; and the occupation number of each such mode is
$<m_k>\sim 1$. These details about the generic state will be important to us later.

\subsection{Entropy for the three charge state}

Let us now ask what happens if we add in the third charge, which will be NS5 branes if the first two charges are NS1-P.
We will build a hand-waving picture for the 3-charge bound state which will be enough for our present purposes; a more systematic derivation of these properties can however be given by using an `orbifold CFT' to describe the bound state.

Suppose we have only one NS5 brane. Since the NS1 brane lies along the NS5 and is bound to the NS5, we can imagine that the NS1 can vibrate inside the plane of the the NS5 but not `come out' of that plane. The momentum P will still be carried by traveling waves along the NS1, but now only four directions of vibration are allowed -- the ones inside the NS5 and transverse to the NS1. Thus $f_B$ in (\ref{five}) is 4 instead of 8. The three charge bound state is supersymmetric, so we should have 4 fermionic excitation modes as well. Then 
\be
c=f_B+{1\over 2} f_F=4+2=6
\ee
 But the rest of the computation can be done as for the two charge case. Using (\ref{sixqq}) we have
 \be
 S=[{c\pi L_T E\over 3}]^{1\over 2}=2\pi\sqrt{n_1n_p}
 \ee
 Since the three charges can be permuted among each other by duality, we expect a permutation symmetric result. Since we have taken $n_5=1$ we can write
 \be
 S=2\pi\sqrt{n_1n_5n_p}
 \ee
To understand the general case of $n_5>1$ we must get some understanding of  why the winding number $n_1$ becomes effectively $n_1n_5$ when we have $n_5$ 5-branes in the system. To do this, recall that by dualities we have the map 
\be
 NS1 (n_1) ~~ P (n_p) ~ \leftrightarrow ~ NS5 (n_1) ~~ NS1 (n_p)
 \ee
  So let us first look at NS1-P. Suppose the NS1 wraps only {\it once} around the $S^1$. The $n_p$ units of momentum are partitioned among different harmonics, with the momentum of the excitations coming in multiples of $1/R$. Now suppose the NS1 is wound $n_1>1$ times around the $S^1$. The total length of the `multiwound' string  is now $2\pi R n_1$ and the momentum now comes in multiples of 
  \be
  \Delta p=1/(n_1 R)
  \ee
   (The total momentum $n_p/R$ must still be an integer multiple of $1/R$, since this quantization must be true for any system living on the $S^1$ of radius $R$ \cite{dmfrac}.) We therefore have $n_1n_p$ units of `fractional' strength $\Delta p$ that we can partition in  different ways to get the allowed states of the system. 

Now consider the NS5-NS1 system obtained after duality. If there is only one NS5 (i.e. $n_1=1$) then we just have $n_p$ NS1 branes bound to it. Noting how different states were obtained in the NS1-P picture we expect that we can count different states  by partitioning this number $n_p$ in different ways. We can picture this by saying that the NS1 strings live in the NS5, but can be joined up to make `multiwound' strings in different ways. Thus we can have $n_p$ separate singly wound loops, or one loop wound $n_p$ times, or any other combination such that the total winding is $n_p$:
\be
\sum_i m_i k_i  = n_p
\ee
where we have $m_i$ strings with winding number $k_i$. 

 If on the other hand we have many NS5 branes ($n_1>1$) then duality indicates that the NS1 breaks up into `fractional' NS1 branes, so that there are $n_1n_p$ strands in all. These latter strands can now be grouped together in various ways so that the number of possible states is given by partitions of $n_1n_p$
 \be
 \sum_i m_i k_i=n_1n_p
 \label{six}
 \ee
In fact we should be able to reproduce the entropy (\ref{three}) by counting such partitions. Let us call each `multiwound' strand in the above sum a `component string'. The only other fact that we need to know about these component strings is that they have 4 fermion zero modes coming from left movers and 4 from right movers; this can be established by a more detailed treatment of the bound states using the `orbifold CFT'. Upon quantization we get two `raising operators' and two `lowering operators' for each of the left and right sides. Starting with a ground state (annihilated by all lowering operators) 
 we can choose to apply or not apply each of the 4 possible raising operators, so we get $2^4=16$ possible ground states of the component string. Applying an even number of raising operators gives a bosonic state while applying an odd number gives a fermionic state. Each component string (with a given winding number $k$) has therefore 8 bosonic states and 8 fermionic states. 
 
 The count of possible states of the NS5-NS1 system is now just like the count for the NS1-P system, done by method (b). If we partition the number $n_1n_p$ as in (\ref{six}) and there are 8 bosonic and 8 fermionic states for each member in a partition, then the total number of states ${\cal N}$ will be given by
 \be
 \ln[{\cal N}]=2\sqrt{2}\pi\sqrt{n_1n_p}
 \ee
 
 With this understanding, let us return to the 3-charge system we were studying. We have $n_5$ NS5 branes and $n_1$ NS1 branes. The bound state of these two kinds of branes will generate an `effective string' which has total winding number
 $n_1n_5$ \cite{maldasuss}. This effective string can give rise to many states where the `component strings' of the state have windings $k_i$ with
 \be
 \sum m_i k_i =n_1n_5
 \label{seven}
 \ee
 
We will later use  a special subclass of states where all the component strings have the same winding $k$; we will also let each component string have the same fermion zero modes. Then the number of component strings is
\be
m={n_1n_5\over k}
\label{teight}
\ee
In the above set, one extreme case is where all component strings are singly wound
\be
k=1, ~~~m=n_1n_5
\label{eight}
\ee
The other extreme is where there is only one component string
\be
k=n_1n_5, ~~~m=1
\label{nine}
\ee

Let us now add the momentum charge $P$ to the system. We can take the NS1-NS5 bound state to be in any of the configurations (\ref{seven}), and the $n_p$ units of momentum can be distributed on the different component strings
in an arbitrary way. All the states arising in this way will be microstates of the NS1-NS5-P system, and should be counted towards the entropy. But one can see that at least for small values of $n_p$ we get a larger contribution from the case if we have only a small number of component strings, each having a large $k_i$. To see this consider the case where $n_p=1$. First consider the extreme case (\ref{eight}). Since each component string is singly wound ($k=1$) there is no `fractionation', and we just place one unit of momentum on any one of the component strings. Further since all the component strings are alike (we chose all component strings to have the same zero modes) we do not get different states by exciting different component strings. Instead we have a state of the form
\be
|\Psi\rangle={1\over \sqrt{n_1n_5}}[({\rm component ~string~ 1 ~excited})~+~\dots ~+~({\rm component ~string~ n_1n_5 ~excited})]
\ee
The momentum mode can be in 4 bosonic states and 4 fermionic states, so we just get 8 states for the system. 

Now consider the other extreme (\ref{nine}). There is only one component string, but since it has winding $w=n_1n_5$ the one unit of momentum becomes an excitation at level $n_1n_5$ on the component string. The number of states is then given by partitioning this level into among different harmonics, and we get for the number of states
\be
{\cal N}\sim e^{2\pi\sqrt{c\over 6}\sqrt{n_1n_5}}=e^{2\pi\sqrt{n_1n_5}}
\ee
where we have used $c=6$ since we have 4 bosons and 4 fermions. This is much larger than the number of states obtained for the case $k_i=1$.

The leading order entropy for NS1-NS5-P can  be obtained by letting the NS1-NS5 be in the bound state
(\ref{nine}) and ignoring other possibilities. We put the $n_p$ units of momentum on this single long component string, getting an effective level of excitation $n_1n_5n_p$ and an entropy
\be
S_{micro}=\ln [{\cal N}] = 2\pi \sqrt{n_1n_5n_p}
\label{ten}
\ee 

We now observe that the microscopic entropy (\ref{ten}) agrees exactly with the Bekenstein entropy (\ref{threeqp}).

This is a remarkable result, first obtained by Strominger and Vafa \cite{sv} for a slightly different system. They took the compactification $M_{4,1}\times S^1\times K3$ (i.e. the $T^4$ was replaced by K3). The case with $T^4$ was done soon thereafter by Callan and Maldacena \cite{callanmalda}. 

So far we have looked at BPS states, which have the minimum energy possible for their charge. Thus we had taken the momentum modes to run along one direction of the effective string (we can call them left movers). By contrast if we add some momentum modes which move in the {\it other} direction along the effective string (right movers) then we would {\it increase} the energy but {\it decrease} the momentum charge $P$. For a small number $\bar n_p$ of right movers we get  a {\it near-extremal} state. If the component string is long ($n_1, n_5$ large) but $ n_p, \bar n_p$ are small by comparison  then we can view the momentum modes as forming a {\it dilute gas} of excitations on the component string. We can then ignore the interactions between the left and right moving excitations. The entropy is then given by the sum of the entropies of the left and right movers, and we get
\be
S_{micro}=2\pi\sqrt{n_1n_5n_p}+2\pi\sqrt{n_1n_5\bar n_p}
\ee
In \cite{callanmalda} the above microscopic entropy was computed and compared to the Bekenstein entropy $S_{Bek}$ for the near extremal hole. Again an {\it exact } agreement was found between the microscopic and Bekenstein entropies.   

A left moving excitation can collide with a right moving one, whereupon the energy contained in them can leave
the D1-D5 bound state and get radiated away as a massless quantum of supergravity. The spin dependence and radiation rates of this emission process agree {\it exactly} with the low energy Hawking radiation from the corresponding black hole \cite{dm}.

\section{Constructing the microstates}

We have seen that  string theory gives us a count of microstates which agrees with the Bekenstein entropy. But to solve the information problem we need to know  what these microstates {\it look} like. We want to understand the structure of states in the coupling domain where we get the black hole. This is in contrast to a  count at $g=0$ which can give us the correct {\it number} of states (since BPS states do not shift under changes of $g$) but will not tell us what the inside of a black hole looks like. 

At this stage we already notice a puzzling fact. For the three charge case we found $S_{micro}=S_{Bek}=2\pi\sqrt{n_1n_5n_p}$. But suppose we keep only two of the charges, setting say $n_5=0$. Then the Bekenstein entropy $S_{Bek}$  becomes zero; this is why we  had to take three charges to get a good black hole. But the microscopic entropy for two charges NS1-P was $S_{micro}=2\pi\sqrt{2}\sqrt{n_1n_p}$, which is nonzero. 

One might say that the 2-charge case is just not a system that gives a good black hole, and should be disregarded in our investigation of black holes. But this would be strange, since on the microscopic side the entropy of the 2-charge system arose in a very similar way to that for the three charge system; in each case we partitioned among harmonics the momentum 
on a string or `effective string'. We would therefore like to take a closer look at the gravity side of the problem for the case of two charges. 

We get the metric for NS1-P by setting to zero the $Q_5$ charge in  (\ref{tenp}). With a slight change of notation we write the metric as ($u=t+y, ~v=t-y$)
 \bea
ds^2_{string}&=&H[-dudv+Kdv^2]+\sum_{i=1}^4 dx_idx_i+\sum_{a=1}^4 dz_adz_a\n \\
B_{uv}&=&-{1\over 2}[H-1]\n \\
e^{2\phi}&=&H\n \\
H^{-1}&=&1+{Q_1\over r^2}, ~~\qquad K={Q_p\over r^2}
\label{naive}
\eea
We will call this metric the {\it naive} metric for NS1-P. This is because we will later argue that this metric is not produced by any configuration of NS1, P charges. It is a solution of the low energy supergravity equations away from $r=0$, but just because we can write such a solution does not mean that the singularity at $r=0$ will be an allowed one in the full string theory. 

What then are the singularities that {\it are} allowed? If we start with flat space, then string theory tells us that excitations around flat space are described by configurations of various fundamental objects of the theory; in particular, the fundamental string. We can wrap this string around  a circle like the $S^1$ in our compactification. We have also seen that we can wrap this string $n_1$ times around the $S^1$ forming a bound state. For $n_1$ large this configuration will generate the solution which has only NS1 charge
 \bea
 ds^2_{string}&=&H[-dudv]+\sum_{i=1}^4 dx_idx_i+\sum_{a=1}^4 dz_adz_a\n \\
B_{uv}&=&-{1\over 2}[H-1]\n \\
e^{2\phi}&=&H\n \\
H^{-1}&=&1+{Q_1\over r^2}
\label{f1}
\eea
This solution is also singular at $r=0$, but this is a singularity that we must accept since the geometry was generated by a source that exists in the theory. One may first take the limit $g\r 0$ and get the string wrapped $n_1$ times around $S^1$ in flat space. Then we can increase $g$ to a nonzero value, noting that we can track the state under the change since it is a BPS state. If $n_1$ is large and we are not too close to $r=0$ then (\ref{f1}) will be a good description of the solution corresponding to the bound state of $n_1$ units of NS1 charge. 

Now let us ask what happens when we add P charge. We have already seen that in the bound state NS1-P the momentum P will be carried as traveling waves on the `multiwound' NS1. Here we come to the most critical point of our analysis: {\it There are no longitudinal vibration modes of the fundamental string NS1}. Thus all the momentum must be carried by transverse vibrations. But this means that the string must bend away from its central axis in order to carry the momentum, so it will not be confined to the location $r=0$ in the transverse space. We will shortly find the correct solutions for NS1-P, but we can already see that the solution (\ref{naive}) may be incorrect since it requires the NS1-P source to be at a point $r=0$ in the transverse space.

The NS1 string has many strands since it is multiwound. When carrying a generic traveling wave these strands will separate from each other. We have to find the metric created by these strands.  Consider the bosonic excitations, and for the moment restrict attention to the 4 that give bending in the noncompact directions $x_i$. The wave carried by the NS1 is then described by a transverse displacement profile $\vec F(v)$, where $v=t-y$. The metric for a single strand of the string carrying such a wave is known \cite{wave}
 \bea
ds^2_{string}&=&H[-dudv+Kdv^2+2A_i dx_i dv]+\sum_{i=1}^4 dx_idx_i+\sum_{a=1}^4 dz_adz_a\nn
B_{uv}&=&-{1\over 2}[H-1], ~~\qquad B_{vi}=HA_i\nn
e^{2\phi}&=&H\nn
H^{-1}(\vec x ,y,t)&=&1+{Q_1\over |\vec x-\vec F(t-y)|^2}\nn
K(\vec x ,y,t)&=&{Q_1|\dot{\vec F}(t-y)|^2\over |\vec x-\vec F(t-y)|^2}\nn
A_i(\vec x ,y,t)&=&-{Q_1\dot F_i(t-y)\over |\vec x-\vec F(t-y)|^2}
\label{fpsingle}
\eea
Now suppose that we have many strands of the NS1 string,  carrying different vibration profiles $\vec F^{(s)}(t-y)$. 
While the vibration profiles are different, the strands all carry momentum in the same direction $y$. In this case the strands are mutually BPS and the metric of all the strands can be obtained by superposing the harmonic functions arising in the solutions for the individual strands. Thus we get
 \bea
ds^2_{string}&=&H[-dudv+Kdv^2+2A_i dx_i dv]+\sum_{i=1}^4 dx_idx_i+\sum_{a=1}^4 dz_adz_a\n \\
B_{uv}&=&-{1\over 2}[H-1], ~~\qquad B_{vi}=HA_i\n \\
e^{2\phi}&=&H\n \\
H^{-1}(\vec x, y,t)&=&1+\sum_s{Q_1^{(s)}\over |\vec x-\vec F^{(s)}(t-y)|^2}\n \\
K(\vec x, y,t)&=&\sum_s{Q_1^{(s)}|\dot{\vec F}^{(s)}(t-y)|^2\over |\vec x-\vec F^{(s)}(t-y)|^2}\n \\
A_i(\vec x,y,t)&=&-\sum_s{Q_1^{(s)}\dot F^{(s)}_i(t-y)\over |\vec x-\vec F^{(s)}(t-y)|^2}
\label{fpmultiple}
\eea

Now consider the string that we actually have in our problem. 
 We can open up the multiwound string by going to the $n_1$ fold cover of $S^1$. Then the string is described by the profile $\vec F(t-y)$, with $0\le y<2\pi R n_1$.  The part of the string in the range $0\le y<2\pi R$ gives one strand in the actual space, the part in the range $2\pi R\le y<4\pi R$ gives another strand, and so on. These different strands do not lie on top of each other in general, so we have a many strand situation as in (\ref{fpmultiple}) above. But note that the end of one strand is at the same position as the start of the next strand, so the strands are not completely independent of each other. In any case all strands are given once we give the profile function $\vec F(v)$.
 
 The above solution has a sum over strands that looks difficult to carry out in practice. But now we note that there is a simplification in the `black hole' limit which is defined by
 \be
 n_1, n_p\r \infty
 \label{fiftqq}
 \ee
 while the moduli like $g, R, V$ are held fixed. We have called this limit the black hole limit for the following reason.
 As we increase the number of quanta $n_i$ in a bound state, the system will in general change its behavior and properties. In the limit  $n_i\r\infty$ we expect that there will be a certain set of properties that will govern the system, and these are the properties that will be the universal ones that characterize  large black holes (assuming that the chosen charges do form a black hole).
 
  The total length of the NS1 multiwound string is $2\pi n_1 R$. From (\ref{thir}) we see that the generic vibration profile has harmonics of order $k\sim \sqrt{n_1n_p}$ on this string, so the wavelength of the vibration is 
 \be
 \lambda\sim {2\pi Rn_1\over \sqrt{n_1n_p}}\sim {\sqrt{n_1\over n_p}}R
 \label{fift}
 \ee
 We will see shortly that the generic state of the string is not well described by a classical geometry, so we will first take some limits to get good classical solutions, and use the results to estimate the `size' of the generic `fuzzball'. Let us take a state where the typical wavenumber is much smaller than the value (\ref{thir})
 \be
 {k\over \sqrt{n_1n_p}}\equiv \alpha \ll1
 \ee
Then the wavelength of the vibrations is much longer than the length of the compactification circle
\be
\lambda={2\pi R n_1\over k}={2\pi R \over \alpha}\sqrt{n_1\over n_p}\gg2\pi R
\ee
where we have assumed that $n_1, n_p$ are of the same order. 

When executing its vibration the string will move in the transverse space across a coordinate distance
\be
\Delta x\sim |\dot{\vec F}|\lambda
\ee
But the distance between neighboring strands of the string will be
\be
\delta x=|\dot{\vec F}|(2\pi R)
\ee
We thus see that
\be
{\delta x\over \Delta x}\sim \sqrt{n_p\over n_1}~\alpha\ll1
\ee

We can therefore look at the metric at points that are not too close to any one of the strands, but that are still in the general region occupied
by the vibrating string
\be
|\vec x-\vec F(v)|\gg\delta x
\ee
(We will later address the nature of the solution as we approach the strands, after dualizing to the D1-D5 picture.) In this case neighboring strands give very similar contributions to the harmonic functions in (\ref{fpmultiple}), and we may replace the sum by an integral
\be
\sum_{s=1}^{n_1} \r \int _{s=0}^{n_1} ds = \int_{y=0}^{2\pi R n_1}{ds\over dy} dy
\ee
Since the length of the compacification circle is $2\pi R$ we have
\be
{ds\over dy}={1\over 2\pi R}
\ee
Also, since the vibration profile is a function of $v=t-y$ we can replace the integral over $y$ by an integral over $v$. Thus we have
\be
\sum_{s=1}^{n_1}\r {1\over 2\pi R}\int_{v=0}^{L_T }dv
\ee
where 
\be
L_T=2\pi R n_1
\label{fseven}
\ee
is the total range of the $y$ coordinate on the multiwound string. Finally, note that
\be
Q_1^{(i)}={Q_1\over n_1}
\ee
We can then write the NS1-P solution as
 \bea
ds^2_{string}&=&H[-dudv+Kdv^2+2A_i dx_i dv]+\sum_{i=1}^4 dx_idx_i+\sum_{a=1}^4 dz_adz_a\nn
B_{uv}&=&-{1\over 2}[H-1], ~~\qquad B_{vi}=HA_i\nn
e^{2\phi}&=&H
\label{ttsix}
\eea
where
\bea
H^{-1}&=&1+{Q_1\over L_T}\int_0^{L_T}\! {dv\over |\vec x-\vec F(v)|^2}\\
K&=&{Q_1\over
L_T}\int_0^{L_T}\! {dv (\dot
F(v))^2\over |\vec x-\vec F(v)|^2}\\
A_i&=&-{Q_1\over L_T}\int_0^{L_T}\! {dv\dot F_i(v)\over |\vec x-\vec F(v)|^2}
\label{functionsq}
\eea

\subsection{Obtaining the D1-D5 geometries}

From (\ref{twop}) we see  that we can perform S,T dualities to map the above NS1-P solutions to D1-D5 solutions.
For a detailed presentation of the steps (for a specific $\vec F(v)$) see \cite{lm3}. The computations are straightforward, except for one step where we need to perform an electric-magnetic duality. Recall that under T-duality a Ramond-Ramond gauge field form $C^{(p)} $
can change to a higher form $C^{(p+1)} $ or to a lower form $C^{(p-1)} $. We may therefore find ourselves with $C^{(2)}$ and $C^{(6)}$ at the same time in the solution. The former gives $F^{(3)}$ while the latter gives $F^{(7)}$. We should convert the $F^{(7)}$ to $F^{(3)}$ by taking the dual, so that the solution is completely described using only $C^{(2)}$.  Finding $F^{(3)}$ is straightforward, but it takes some inspection to find a $C^{(2)}$ which will give this $F^{(3)}$. 

 Note that we have chosen to write the classical solutions in a way where $\phi$ goes to zero at infinity, so that the true dilaton $\hat\phi$ is given by
\be
e^{\hat\phi}=ge^\phi
\ee
The dualities change the values of the moduli describing the solution. Recall that the $T^4$ directions are $x^6, x^7, x^8, x^9$, while the $S^1$ direction is $y\equiv x^5$. We keep track of (i) the coupling $g$ (ii) the value of the scale $Q_1$ which occurred in the harmonic function for the NS1-P geometry (iii) the radius $R$ of the $x^5$ circle (iv) the radius $R_6$ of the $x^6$ circle, and (v) the volume $(2\pi)^4 V$ of $T^4$.  We can start with NS1-P and reach NS5-NS1, which gives
(here we set $\alpha'=1$ for compactness)
\be\label{DualParam}
\left(\begin{array}{c}
g\\Q_1\\R\\R_6\\V
\end{array}\right)
\stackrel{\textstyle S}{\rightarrow}
\left(\begin{array}{c}
1/g\\Q_1/{g}\\R/\sqrt{g}\\R_6/\sqrt{g}\\V/g^2
\end{array}\right)
\stackrel{\textstyle T6789}{\rightarrow}
\left(\begin{array}{c}
g/V\\Q_1/{g}\\R/\sqrt{g}\\\sqrt{g}/R_6\\g^2/V
\end{array}\right)
\stackrel{\textstyle S}{\rightarrow}
\left(\begin{array}{c}
V/g\\Q_1{V}/g^2\\R\sqrt{V}/g\\\sqrt{V}/R_6\\V
\end{array}\right)
\stackrel{\textstyle T56}{\rightarrow}
\left(\begin{array}{c}
R_6/R\\Q_1{V}/g^2\\g/(R\sqrt{V})\\R_6/\sqrt{V}\\R_6^2
\end{array}\right)
\ee
A final S-duality takes the NS5-NS1 to D5-D1
\be
\left(\begin{array}{c}
R_6/R\\Q_1{V}/g^2\\g/(R\sqrt{V})\\R_6/\sqrt{V}\\R_6^2 
\end{array}\right)
\stackrel{\textstyle S}{\rightarrow}
\left(\begin{array}{c}
R/R_6\\Q_1 V {R}/( g^2 {R_6})\\g/(\sqrt{RR_6V})\\ \sqrt{R_6R/V}\\R^2
\end{array}\right)
\equiv
\left(\begin{array}{c}
g'\\Q_5'\\R'\\R'_6\\V'
\end{array}\right)
\label{eightt}
\ee
where at the last step we have noted that the $Q_1$ charge in NS1-P becomes the D5 charge $Q'_5$ in D5-D1. 
We will also choose coordinates at each stage so that the metric goes to $\eta_{AB}$ at infinity. Since we are writing the string metric, this convention is not affected by T-dualities, but when we perform an S-duality we need to re-scale the coordinates to keep the metric $\eta_{AB}$.  In the NS1-P solution the harmonic function generated by the NS1 branes is (for large $r$) 
\be
H^{-1}\approx 1+{Q_1\over r^2}
\label{ninet}
\ee
After we reach the D1-D5 system by dualities the corresponding harmonic function will behave as
\be
H^{-1}\approx 1+{Q'_5\over r^2}
\label{fsix}
\ee
where from (\ref{eightt}) we see that
\be
Q'_5=\mu^2 Q_1
\ee
with
\be
\mu^2={VR\over g^2R_6}
\label{fone}
\ee
Note that $Q_1, Q'_5$ have units of $(length)^2$. Thus all lengths get scaled by a factor $\mu$ after the dualities. Note that
\be
Q'_5=\mu^2Q_1=\mu^2{g^2n_1\over V}=g'n_1
\label{feight}
\ee
which is the correct parameter to appear in the harmonic function (\ref{fsix}) created by the D5 branes.

With all this, for D5-D1 we get the solutions \cite{lm4}
\be
ds^2_{string}=\sqrt{H\over 1+K}[-(dt-A_i dx^i)^2+(dy+B_i dx^i)^2]+\sqrt{1+K\over
H}dx_idx_i+\sqrt{H(1+K)}dz_adz_a
\label{qsix}
\ee
where the harmonic functions are
\bea
H^{-1}&=&1+{\mu^2Q_1\over \mu L_T}\int_0^{\mu L_T} {dv\over |\vec x-\mu\vec F(v)|^2}\nn
K&=&{\mu^2Q_1\over
\mu L_T}\int_0^{\mu L_T} {dv (\mu^2\dot
 F(v))^2\over |\vec x-\mu\vec F(v)|^2},\nonumber\\
A_i&=&-{\mu^2Q_1\over \mu L_T}\int_0^{\mu L_T} {dv~\mu\dot F_i(v)\over |\vec x-\mu\vec F(v)|^2}
\label{functionsqq}
\eea
Here $B_i$ is given by
\be
dB=-*_4dA
\label{vone}
\ee
and $*_4$ is the duality operation in the 4-d transverse  space
$x_1\dots
x_4$ using the flat metric $dx_idx_i$.

By contrast the `naive' geometry which one would write for D1-D5 is
\be
ds^2_{naive}={1\over \sqrt{(1+{Q'_1\over r^2})(1+{Q'_5\over
r^2})}}[-dt^2+dy^2]+\sqrt{(1+{Q'_1\over r^2})(1+{Q'_5\over
r^2})}dx_idx_i+\sqrt{{1+{Q'_1\over r^2}\over 1+{Q'_5\over r^2}}}dz_adz_a
\label{d1d5naive}
\ee

\section{A special example}

The above general solution looks rather complicated. To get a feeling for the nature of these D1-D5 solutions let us start by examining in detail a simple case. Start with the NS1-P solution which has the following vibration profile for the NS1 string
\be
F_1=\hat a\cos\omega v,\quad F_2=\hat a\sin\omega v, \quad F_3=F_4=0
\label{yyonePrime}
\ee
where $\hat a$ is a constant.
This makes the NS1 swing in a uniform helix in the $x_1-x_2$ plane. Choose
\be
\omega=\frac{1}{n_1R}
\ee
This makes the NS1 have just one turn of the helix in the covering space. Thus all the energy has been put in the lowest harmonic on the string.

We then find
\be
H^{-1}=1+{Q_1\over 2\pi}\int_0^{2\pi} {d\xi\over
(x_1-\hat a\cos\xi)^2+(x_2-\hat a\sin\xi)^2+x_3^2+x_4^2}
\ee
To compute the integral we introduce polar coordinates in the $\vec x $ space
\bea\label{EpolarMap}
x_1&=&{\tilde r} \sin{\tilde \theta} \cos{\tildr\phi}, ~~~\qquad x_2={\tilde r}
\sin{\tilde\theta} \sin{\tildr\phi},\nonumber \\
x_3&=&{\tilde r} \cos{\tilde\theta} \cos{\tildr\psi}, ~~\qquad x_4={\tilde r}
\cos{\tilde\theta} \sin{\tildr\psi}
\label{etwo}
\eea
Then we find
\be
H^{-1}=1+{Q_1\over
\sqrt{(\tilde r^2+\hat a^2)^2-4 \hat a^2\tilde r^2\sin^2\tilde\theta}}
\ee

The above expression simplifies if we
change from $\tilde r, \tilde\theta$ to coordinates $r,\theta$:
\bea
{\tilde r}&=& \sqrt{r^2+\hat a^2\sin^2\theta}, ~~\qquad \cos{\tilde\theta}
={r\cos\theta\over \sqrt{r^2+\hat a^2\sin^2\theta}}
\label{ethree}
\eea
(${\tildr\phi}$ and ${\tildr\psi}$ remain unchanged). Then we get
\be
H^{-1}=1+{Q_1\over r^2+\hat a^2\cos^2\theta}
\ee
Similarly we get
\be
K={\hat a^2\over n_1^2 R^2}~{Q_1\over (r^2+\hat a^2\cos^2\theta)}
\ee
With a little algebra we also find
\bea
A_{x_1}&=&{Q_1\hat a\over 2\pi R n_1}\int_0^{2\pi}{d\xi \sin\xi\over (x_1-\hat a\cos\xi)^2+(x_2-\hat a\sin\xi)^2+x_3^2+x_4^2}\nn
&=&{Q_1\hat a\over 2\pi R n_1}\int_0^{2\pi}{d\xi \sin\xi\over (\tilde r^2+\hat a^2-2\tilde r \hat a\sin\tilde\theta\cos(\xi-\tilde\phi))}
\nn
&=&{Q_1\hat a^2\over R n_1}\sin\tilde\phi {\sin\theta\over (r^2+a^2\cos^2\theta)}{1\over \sqrt{r^2+a^2}}
\eea
\bea
A_{x_2}&=&-{Q_1\hat a^2\over R n_1}\cos\tilde\phi {\sin\theta\over (r^2+a^2\cos^2\theta)}{1\over \sqrt{r^2+a^2}}
\eea
\be
A_{x_3}=0, \qquad A_{x_4}=0
\ee
We can write this in polar coordinates
\bea
A_{\tilde\phi}&=&A_{x_1}{\p x_1\over \p\tilde\phi}+A_{x_2}{\p x_2\over \p\tilde\phi}\nn
&=&-{Q_1\hat a^2\over R n_1}{\sin^2\theta\over (r^2+a^2\cos^2\theta)}
\eea
We can now substitute these functions in (\ref{ttsix}) to get the solution for the NS1-P system for the choice of profile (\ref{yyonePrime}).

Let us now get the corresponding D1-D5 solution. Recall that all lengths scale up by a factor $\mu$ given through (\ref{fone}).
The transverse displacement profile $\vec F$  has units of length, and so scales up by the factor $\mu$. We define
\be
a\equiv \mu \hat a 
\ee
so that
\be
\mu F_1= a\cos\omega v,\quad \mu F_2= a\sin\omega v, \quad F_3=F_4=0
\label{yyonePrimep}
\ee
Let
\be
f=r^2+a^2\cos^2\theta
\ee
The NS1 charge becomes the D5 charge after dualities, and corresponding harmonic function becomes
\be
H'^{-1}=1+{Q'_5\over f}
\ee
The harmonic function for momentum P was
\be
K={Q_1\hat a^2\over n_1^2R^2}{1\over (r^2+\hat a^2 \cos^2\theta)}\equiv {Q_p\over (r^2+\hat a^2 \cos^2\theta)}
\ee
After dualities $K$ will change to the harmonic function generated by D1 branes. Performing the change of scale (\ref{fone})
we find
\be
K'=\mu^2 {Q_p\over f}\equiv{Q'_1\over f}
\ee
Using the value of $Q_1$ from (\ref{sixt}) we observe that
\be
a={\sqrt{Q'_1Q'_5}\over R'}
\label{ftwo}
\ee
where $R'$ is the radius of the $y$ circle after dualities (given in (\ref{eightt})). 

To finish writing the D1-D5 solution we also need the functions $B_i$ defined through (\ref{vone}). In the coordinates
$r, \theta, \tilde\phi\equiv\phi, \tilde\psi\equiv\psi$ we have
\be
A_\phi=-{a\sqrt{Q'_1Q'_5}}{\sin^2\theta\over f}
\ee
We can check that the dual form is
\be
B_\psi=-{a\sqrt{Q'_1Q'_5}}{\cos^2\theta\over f}
\ee
To check this, note that the flat 4-D metric in our coordinates is
\be
dx_idx_i={f\over r^2+a^2}dr^2+fd\theta^2+(r^2+a^2)\sin^2\theta d\phi^2+r^2\cos^2\theta d\psi^2
\ee
We also have
\be
\epsilon_{r\theta\phi\psi}=\sqrt{g}=f r\sin\theta\cos\theta
\ee
We then find
\be
F_{r\psi}=\partial_r B_\psi={a\sqrt{Q'_1Q'_5}}{2r\cos^2\theta\over f^2}=-\epsilon_{r\psi\theta\phi}g^{\theta\theta}g^{\phi\phi}[\partial_\theta A_\phi]=-(*dA)_{r\psi}
\ee
\be
F_{\theta\psi}=\partial_\theta B_\psi={a\sqrt{Q'_1Q'_5}}{r^2\sin(2\theta)\over f^2}=-\epsilon_{\theta\psi r\phi}g^{rr}g^{\phi\phi}[\partial_r A_\phi]=-(*dA)_{\theta\psi}
\ee
verifying (\ref{vone}).

Putting all this in (\ref{qsix}) we find the D1-D5 (string) metric for the profile (\ref{yyonePrime})
\bea\label{MaldToCompare}
d{s}^2&=&-\frac{1}{h}(d{t}^2-d{ y}^2)+
hf\left(d\theta^2+\frac{d{r}^2}{{r}^2+a^2}\right)
-\frac{2a\sqrt{Q'_1 Q'_5}}{hf}\left(\cos^2\theta d{ y}d\psi+
\sin^2\theta d{ t}d\phi\right)\nonumber\\
&+&h\left[
\left({ r}^2+\frac{a^2Q'_1Q'_5\cos^2\theta}{h^2f^2}\right)
\cos^2\theta d\psi^2+
\left({ r}^2+a^2-\frac{a^2Q'_1Q'_5\sin^2\theta}{h^2f^2}\right)
\sin^2\theta d\phi^2\right]\nonumber \\
&+&\sqrt{Q'_1+f\over Q'_5+f}~dz_adz_a
\eea
where
\be\label{defFHProp}
f={r}^2+a^2\cos^2\theta,\qquad
h=\left[\left(1+\frac{Q'_1}{f}\right)\left(1+\frac{Q'_5}{f}\right)\right]^{1/2}
\ee

At large $r$ this metric goes over to flat space. Let us consider the opposite limit  $r\ll
(Q'_1Q'_5)^{1/4}$ (we write $r'=r/a$):
\bea
ds^2&=&-({r'}^2+1)\frac{a^2dt^2}{\sqrt{Q'_1Q'_5}}+{r'}^2
\frac{a^2dy^2}{\sqrt{Q'_1Q'_5}}+
\sqrt{Q'_1Q'_5}\frac{d{r'}^2}{{r'}^2+1}\nonumber\\
&+&\sqrt{Q_1Q_5}\left[d\theta^2+\cos^2\theta \left(d{\psi}-
\frac{ady}{\sqrt{Q'_1Q'_5}}\right)^2+
\sin^2\theta \left(d{\phi}-\frac{adt}{\sqrt{Q'_1Q'_5}}\right)^2\right]\nn
&+&\sqrt{Q'_1\over Q'_5}dz_adz_a
\label{fthree}
\eea
Let us transform to new angular coordinates
\be
\psi'=\psi-{a\over \sqrt{Q'_1Q'_5}}y, ~~\qquad \phi'=\phi-{a\over \sqrt{Q'_1Q'_5}}t
\ee
Since $\psi,y$ are both periodic coordinates, it is not immediately obvious that the first of these changes makes sense.
The identifications on these coordinates are
\be
(\psi\r \psi+2\pi, ~~y\r y), ~~\qquad (\psi\r\psi, ~~y\r y+2\pi R')
\ee
But note that we have the relation (\ref{ftwo}), which implies that the identifications on the new variables are
\be
(\psi'\r \psi'+2\pi, ~~y\r y), ~~\qquad (\psi'\r \psi'-{a2\pi R'\over \sqrt{Q'_1Q'_5}}=\psi'-2\pi, ~~y\r y+2\pi R')
\ee
so that we do have a consistent lattice of identifications on $\psi',y$. 
The metric (\ref{fthree}) now becomes
\bea
\label{esix}
ds^2&=&\sqrt{Q'_1Q'_5}\left[
-({r'}^2+1)\frac{dt^2}{R^2}+{r'}^2
\frac{dy^2}{R^2}+
\frac{d{r'}^2}{{r'}^2+1}\right]\nonumber\\
&+&\sqrt{Q'_1Q'_5}\left[d\theta^2+\cos^2\theta d{\psi'}^2+
\sin^2\theta d{\phi'}^2\right]+\sqrt{Q'_1\over Q'_5}dz_adz_a
\eea
This is just $AdS_3\times S^3\times T^4$. Thus the full geometry is flat at infinity, has a `throat' type region at smaller $r$
where it approximates the naive geometry (\ref{d1d5naive}), and then instead of a singularity at $r=0$ it ends in a smooth `cap'. This particular geometry, corresponding to the profile (\ref{yyonePrime}), was derived earlier in \cite{bal,mm} by taking limits of general rotating black hole solutions found in \cite{cy}. We have now obtained it by starting with the particular NS1-P profile (\ref{yyonePrime}), and thus we note that it is only one member of the complete family parametrized by $\vec F$. It can be shown that all the metrics of this family have the same qualitative structure as the particular metric that we studied; in particular they have no horizons, and they end in smooth `caps' near $r=0$. We will review the argument for this smoothness below.

\section{Regularity of the general solution}

At first sight it appears that the metrics (\ref{qsix}) have a singularity at points where $\vec x=\mu \vec F(v)$ for some $v$. But it was shown in \cite{lmm} that this is only a coordinate singularity; the geometries are completely smooth. To see this we compute the harmonic functions near such points. The points $\vec x=\mu \vec F(v)$ define a curve in the 4-dimensional Cartesian space spanned by the $x_i$; this space carries the flat metric $dx_idx_i$. Go to a point on the curve given by $v=v_0$. Let $z$ be a coordinate that measures distance along the curve (in the flat metric) and choose spherical polar coordinates $(\rho, \theta, \phi)$ for the 3-plane perpendicular to the curve. Then
\be
z\approx \mu |\dot {\vec F}(v_0)| (v-v_0)
\ee
\be
H^{-1}\approx {Q'_5\over \mu L_T}\int_{-\infty}^\infty {dv\over \rho^2+z^2}={Q'_5\over \mu L_T}{1\over \mu|\dot {\vec F}(v_0)|} \int_{-\infty}^\infty {dz\over \rho^2+z^2}={Q'_5\pi\over \mu^2 L_T|\dot {\vec F}(v_0)|}{1\over \rho}
\ee
\be
K\approx {Q'_5\pi |\dot {\vec F}(v_0)|\over  L_T}{1\over \rho}\,,\quad
A_z\approx -{Q'_5\pi\over \mu L_T}{1\over \rho}
\ee
Let
\be
\tilde Q\equiv {Q'_5\pi\over \mu L_T}
\ee
The field $A_i$ gives $F_{z\rho}=-{\tilde Q/ \rho^2}$, which gives from (\ref{vone}) $B_\phi=-
\tilde Q(1-\cos\theta)$. The $(y, \rho, \theta, \phi)$ part of the metric  becomes
\bea
ds^2&\rightarrow& \sqrt{H\over 1+K}(dy+B_i dx^i)^2+\sqrt{1+K\over
H}(d\rho^2+\rho^2(d\theta^2+\sin^2\theta d\phi^2))\nonumber\\
&\approx&{\rho\over \tilde Q}(dy-{\tilde Q}(1-\cos\theta) d\phi)^2+ {\tilde Q\over \rho}
(d\rho^2+\rho^2(d\theta^2+\sin^2\theta d\phi^2))
\label{vfourt}
\eea
This is the metric near the core of a Kaluza-Klein monopole, and is smooth if the length of the $y$ circle is $4\pi \tilde Q$, which implies
\be
 2\pi R' =4\pi ({Q'_5\pi\over \mu L_T})
 \label{fnine}
\ee
Using (\ref{fseven}),(\ref{eightt}),(\ref{fone}),(\ref{feight}) we see that this relation is exactly satisfied. The change of coordinates
\bea
&&{\tilde r}^2=\rho, ~~
\tilde\theta={\theta\over 2}, ~~\tilde y={y\over 2{\tilde Q}}, ~~\tilde \phi=\phi-{y\over 2{\tilde Q}}
\nonumber\\ 
&&0\le\tilde\theta<{\pi\over 2}, ~~0\le \tilde y<{\pi R'\over {\tilde Q}}=2\pi, ~~0\le\tilde\phi<2\pi
\label{fifty}
\eea
makes manifest the locally $R^4$ form of the metric
\be
ds^2=4{\tilde Q}[d\tilde r^2+\tilde r^2(d\tilde\theta^2+\cos^2\tilde\theta d\tilde y^2+\sin^2\tilde\theta d\tilde\phi^2)]
\ee
For the coordinate change (\ref{fifty}) to be consistent with identifications on the periodic coordinates we need the condition (\ref{fnine}), which we have seen is satisfied.

The $(t,z)$ part of the geometry gives
\bea
ds^2&\rightarrow&-\sqrt{H\over 1+K}(dt-A_zdz)^2+\sqrt{1+K\over
H}dz^2\nonumber\\
&\approx&-{\rho\over \tilde Q}dt^2-2 dtdz\approx -2dtdz
\eea
which is regular. The $T^4$ part gives $\mu|\dot {\vec F}(v_0)| dz_a dz_a$ and is thus regular as well. 

We thus see that the generic D1-D5 geometry is qualitatively similar to the specific case that we studied in the last section. The geometry is flat at infinity, and instead of a singularity at $r=0$ it ends in a smooth `cap'. Different profile functions $\vec F(v)$ give different caps. We thus have an ensemble of solutions rather than just the naive solution (\ref{d1d5naive}). Since different $\vec F(v)$ give different possible states of the NS1-P system, and thus of the dual D1-D5 system, we see explicitly the `hair' which distinguishes different microstates.

\section{`Size' of the bound state}

The most important point that we have seen in the above discussion is that in the NS1-P bound state the NS1 undergoes transverse vibrations that cause its strands to spread out over a nonzero range in the transverse $\vec x$ space. Thus the bound state is not `pointlike'. Exactly how big {\it is} the bound state?

We have obtained good classical solutions by looking at solutions where the wavelength of vibrations $\lambda$ was much longer than the wavelength for the generic solution. To get an estimate of the size of the generic state we will now take our classical solutions and extrapolate  to the domain where $\lambda$ takes its generic value (\ref{fift}). 

The wavelength of vibrations  for the generic state is
\be
\lambda={L_T\over k}\sim {2\pi R n_1\over \sqrt{n_1n_p}}\sim R\sqrt{n_1\over n_p}
\ee
We wish to ask how much the transverse coordinate $\vec x$ changes in the process of oscillation. Thus we set $\Delta y=\lambda$, and find
\be
\Delta x\sim |\dot{\vec F}|\Delta y\sim |\dot{\vec F}|R\sqrt{n_1\over n_p}
\ee
Note that
\be
Q_p \sim Q_1 |\dot{\vec F}|^2
\ee
which gives
\be
\Delta x\sim \sqrt{Q_p\over Q_1}R\sqrt{n_1\over n_p}\sim \sqrt{\alpha'}
\label{ffone}
\ee
where we have used (\ref{sixt}). 

For 
\be
|\vec x|\gg\Delta x
\ee
we have
\be
{1\over |\vec x-\vec F|^2}\approx {1\over |\vec x|^2}
\ee
and the solution becomes approximately the naive geometry (\ref{naive}).

We see that the metric settles down to the naive metric outside a certain ball shaped region $|\vec x|>\sqrt{\alpha'}$. 
Let us now ask an important question: What is the surface area of this ball?

First we compute the area in the 10-D string metric. Note that the metric will settle down to the naive form (\ref{naive}) quite rapidly as we go outside the region occupied by the vibrating string.  The mean wavenumber is $k\sim \sqrt{n_1n_p}$, so there are $\sim \sqrt{n_1n_p}$ oscillations of the string. There is in general no correlation between the directions of oscillation in each wavelength, so the string makes $\sim \sqrt{n_1n_p}$ randomly oriented traverses in the ball that we are investigating. This causes a strong cancellation of the leading moments like dipole, quadrupole ...etc. The surviving moments will be of very high order, an order that will increase with $n_1, n_p$ and which is thus infinite in the classical limit of large charges. 

We must therefore compute the area, in the naive metric (\ref{naive}), of the location $|\vec x| =\sqrt{\alpha'}$. Introduce polar coordinates on the 4-D transverse space
\be
d\vec x\cdot d\vec x=dr^2+r^2 d\Omega_3^2
\ee
At the location $r=\sqrt{\alpha'}$ we get from the angular $S^3$ an area
\be
A_{S^3}\sim \alpha'^{3\over 2}
\label{tsix}
\ee
From the $T^4$ we get an area
\be
A_{T^4}\sim V
\ee
From the $S^1$ we get a length
\be
L_y\sim \sqrt{HK} R \sim \sqrt{Q_p\over Q_1} R
\ee
Thus the area of the 8-D surface bounding the region occupied by the string is given, in the string metric, by
\be
A^{S}\sim \sqrt{Q_p\over Q_1}RV{\alpha'}^{3\over 2}
\ee
The area in Einstein metric will be
\be
A^E=A^{S}e^{-2\phi}
\ee
Note that the dilaton $\phi$ becomes very negative at the surface of interest
\be
e^{-2\phi}=H^{-1}\approx {Q_1\over r^2}\sim {Q_1\over \alpha'}
\label{tfive}
\ee
We thus find
\be
A^E\sim  \sqrt{Q_1Q_p}RV\alpha'^{1\over 2}\sim {g^2\alpha'^4}\sqrt{n_1n_p}
\ee
where we have used (\ref{sixt}).
Now we observe that
\be
{A^E\over 4G_{10}}\sim \sqrt{n_1n_p}\sim S_{micro}
\label{fftwo}
\ee
This is very interesting, since it shows that the surface area of our `fuzzball' region satisfies a Bekenstein type relation
\cite{lm5}.

     \begin{figure}[htbp]
   \begin{center}
   \includegraphics[width=6in]{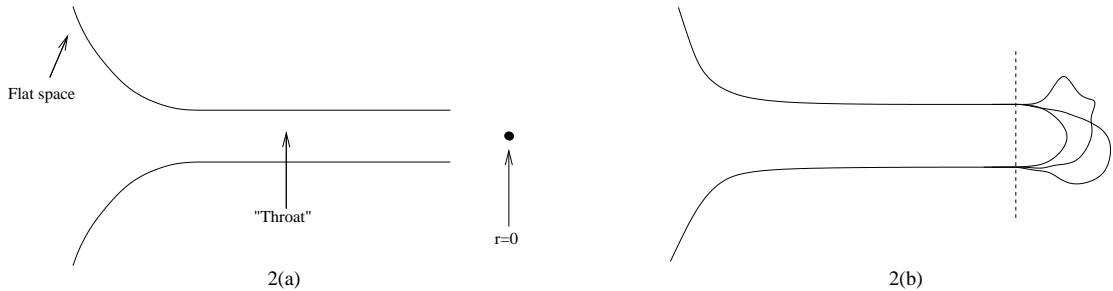}
   \caption{(a) The naive geometry of extremal D1-D5 \quad (b) the actual geometries; the area of the surface denoted by the dashed line reproduces the microscopic entropy.}
   \label{fig2}
   \end{center}
   \end{figure}

\subsection{Nontriviality of the `size'}

One of the key ideas we are pursuing is the following. To make a big black hole we will need to put together many elementary quanta. What is the size of the resulting bound state? One possibility is that this size is always of order planck length $l_p$ or string length $l_s$. In this case we will not be able to avoid the traditional picture of the black hole. Since the horizon radius can be made arbitrarily large, the neighborhood of the horizon will be `empty space' and the matter making the black hole will sit in a small neighborhood of the singularity. But a second possibility is that the size of a bound state {\it increases} with the number of quanta in the bound state
\be
{\cal R}\sim N^\alpha l_p
\label{ttwo}
\ee
where ${\cal R}$ is the radius of the bound state, $N$ is some count of the number of quanta in the state, and the power $\alpha$ depends on what quanta are being bound together. It would be very interesting if in every case we find that
\be
{\cal R}\sim R_H
\label{tthree}
\ee
where $R_H$ is the radius of the horizon that we would find for the classical geometry which has the mass and charge carried by these $N$ quanta. For in that case we would find that we do not get the traditional black hole; rather we get a `fuzzball' which has a size of order the horizon size. Since we do not  get a traditional horizon we do not have  the usual computation of Hawking radiation which gives information loss. The different configurations of the fuzzball will correspond to the $e^{S_{Bek}}$ states expected from the Bekenstein entropy.

For the 1-charge system we saw that the Bekenstein entropy was $S_{Bek}=0$. We also find no nontrivial size for the bound state, so the size remains order $l_p$ or $l_s$, with the exact scale depending perhaps on the choice of probe. This is consistent with the picture we are seeking, but not a nontrivial illustration of the conjecture. But the situation was much more interesting when we moved to the 2-charge case. The microscopic entropy was $S_{micro}=2\sqrt{2}\sqrt{n_1n_p}$. The size of the bound state was such that the area of the boundary satisfied a Bekenstein type relation. We had verified this relation using the 10-D metric, but we can also write it in terms of quantities in the dimensionally reduced 5-D theory
\be
{A_5\over 4 G_5}\sim \sqrt{n_1n_p}
\label{tone}
\ee
We define the 5-D planck length by
\be
l_p^{(5)}\equiv G_5^{1\over 3}
\ee
We also define the radius of the horizon from the area
\be
{\cal R}=[{A_5\over 2\pi ^2}]^{1\over 3}
\ee
The result (\ref{tone}) then translates to
\be
{\cal R}\sim (n_1n_p)^{1\over 6} l^{(5)}_p
\label{tfour}
\ee
Thus for the 2-charge system we find a manifestation of the conjectured relations (\ref{ttwo}), (\ref{tthree}).

While we see from (\ref{tfour}) that the fuzzball size ${\cal R}$ is much larger then planck length $l_p$, we have not yet compared
${\cal R}$ to the string length $l_s$. From (\ref{tsix}) we see that 
\be
{\cal R}\sim \sqrt{\alpha'}\sim l_s
\label{tseven}
\ee
One might think that this indicates that the fuzzball is really small in some sense; it has just the natural minimum radius set by string theory. But such is not the case. In the NS1-P system that we are looking at $e^\phi$ becomes very small at the fuzzball surface. Thus the string tension becomes very low in planck units; in other words, the string becomes very long and `floppy'. Thus we should interpret (\ref{tseven}) as telling us that string length is very large, not that ${\cal R}$ is very small.  

This may sound more a matter of language rather than physics, but we can make it more precise by looking at the D1-D5 system which is obtained from NS1-P by dualities. It is a general result that the area of a surface $r=const$ (measured in planck units) does not change upon dualities. To see this, note that the Einstein action in D spacetime dimensions scales with the metric as follows
\be
S\sim {1\over G_D}\int d^D x \sqrt{-g}R \sim {[g_{ab}]^{D-2\over 2}\over G_D}
\label{ssone}
\ee
This action must remain unchanged under S,T dualities. The hypersurface at fixed $r$ is $D-2$ dimensional. Under dualities the $D-2$ dimensional area scales as $[g_{ab}]^{D-2\over 2}$. We have $G_D=(l_p^{(D)})^{D-2}$. From the invariance  of (\ref{ssone}) we see that  under dualities the scalings are such that the area of the $D-2$ dimensional fuzzball boundary measured in D-dimensional planck units remains invariant.  This fact remains true whether we use a 10-D description or the dimensionally reduced 5-D one. 

 For the fuzzball boundary in the D1-D5 system we get
\be
{A_{10}\over 4 G_{10}}={A_5\over 4 G_5}\sim \sqrt{n_1n_5}
\ee
(We have re-labeled the charges as $n_1\r n_5, n_p\r n_1$  to give them their appropriate names in the D1-D5 geometry.)
Thus 
\be
{\cal R}_{\rm D1-D5}\sim (n_1n_5)^{1\over 6} l_p^{(5)}
\ee
But this time the dilaton does not become strongly negative near the fuzzball boundary; rather it remains order unity
\be
e^{-2\phi}\approx {Q_5\over Q_1}\sim {n_5 V\over n_1\alpha'^2}
\ee
To find ${\cal R}$ in terms of string length and other moduli we write the 10-D area-entropy relation
\be
{A_{10}\over 4G_{10}}\sim {{\cal R}_{\rm D1-D5}^3 V R\over g^2\alpha'^4} \sim \sqrt{n_1n_5}
\ee
to get
\be
{\cal R}_{\rm D1-D5}\sim [{g^2\alpha'^4\over VR}]^{1\over 3} (n_1n_5)^{1\over 6}
\ee
We thus see that $l_p$ and $l_s$ are of the same order (their ratio does not depend on $n_1, n_5$) while ${\cal R}$ grows much larger than these lengths as the number of quanta in the bound state is increased. 

Thus we see that comparing the bound state size to the string length is not a duality invariant notion, while comparing it to the 5-D planck length is. In planck units the bound state size grows with charges. In string units, it also grows with charges in the D1-D5 duality frame, while it remains $l_s$ in the NS1-NS5 duality frame. The latter fact can be traced to the very small value of $e^\phi$ at the fuzzball boundary, which makes the local string length very large in this case.

Sen \cite{sen} looked at the naive geometry of NS1-P, and noticed that the curvature of the string metric became string scale at $r\sim\sqrt{\alpha'}$, the same location that we have found in (\ref{ffone}) as the boundary of the fuzzball. He then argued that one should place a `stretched horizon' at this location, and then observed that the area of this horizon gave the Bekenstein relation (\ref{fftwo}). But if we look at the D1-D5 system that we obtain  by dualities, then the curvature remains {\it bounded and small} as $r\r 0$. The naive geometry is locally $AdS_3\times S^3\times T^4$ for small $r$, and  the curvature radii for the $AdS_3$ and the $S^3$ are $(Q'_1Q'_5)^{1/4}\gg\sqrt{\alpha'}$. So it does not appear that the criterion used by Sen can be applied in general to locate a `stretched horizon'. What we have seen on the other hand is that the naive geometry is not an accurate description for $r$ smaller than a certain value; the interior of this region is different for different states, and the boundary of this region satisfies a Bekenstein type relation (\ref{fftwo}). Further, we get the same relation (\ref{fftwo}) in all duality frames.

We have considered the compactification $T^4\times S^1$, but we could also have considered $K3\times S^1$. Suppose we further compactify a circle $\tilde S^1$, thus getting a 2-charge geometry in 3+1 noncompact dimensions. In this case if we include higher derivative corrections in the action then the naive  geometry develops a horizon, and the area of this horizon correctly reproduces the microscopic entropy \cite{dab}. Order of magnitude estimates suggest that a similar effect will happen for the 4+1 dimensional case that we have been working with. 

 What are the {\it actual} geometries for say D1-D5 on $K3\times S^1$?  Recall that in the NS1-P system that we started with the NS1 string had 8 possible directions for transverse vibration. We have considered the vibrations in the noncompact directions $x_i$; a similar analysis can be carried out for those in the compact directions $z_a$ \cite{lmm}. But after dualizing to D1-D5 we note that the the solutions generated by the $x_i$ vibrations are constant on the compact $T^4$, and we can replace the $T^4$ by $K3$ while leaving the rest of the solution unchanged. (The vibrations along compact directions will be different in the $T^4$ and $K3$ cases, but since we are only looking for estimate of the fuzzball size we ignore these states in the present discussion.) In \cite{gm2} it was shown that the higher derivative terms do not affect the `capped' nature of the `actual' geometries. Thus the K3 case is interesting in that it provides a microcosm of the entire black hole problem: There is a naive geometry that has a horizon, and we have `actual' geometries that have no horizons but that differ from each other inside a region of the size of the horizon.  It would be interesting to understand the effect of higher derivative terms on the naive $T^4$ geometry.

\section{A comment on the absence of horizons}

We have seen that the microstates do not have horizons. But the boundary of the region where these states differ does satisfy a Bekenstein type relation (\ref{fftwo}), so in the fuzzball picture  we can call this the `horizon'.  Defining this horizon is somewhat similar to the idea of `coarse-graining' in statistical physics: If we keep only the part of the geometry outside this `horizon' then we have kept that part of each state which is common to all microstates, and discarded the part where the microstates differ. In usual statistical systems entropy appears when we describe a system by a few macroscopic parameters (which all the microstates share) while ignoring the details that differentiate between microstates. 

The idea that microstates may not have horizons appears radical, so let us investigate this further.
For most physical systems entropy is given by $\ln[N]$ where $N$ is the number of microstates that have the same macroscopic parameters in a coarse grained description. For black holes on the other hand we have a Bekenstein entropy given by the area of the horizon. It is generally believed that if we find any horizon in general relativity we should associate an entropy with it; for example there have been attempts to associate an entropy with the de-Sitter horizon.

It would be nice if the entropy of black holes could also be understood using the same statistical ideas that work for other physical systems. Let us consider the BPS black holes that we have been studying. At weak coupling ($g$ small) we have a large degeneracy of microstates, with $S_{micro}=\ln[N]$. As we increase the coupling these will become black holes, but note that we cannot lose any states from the count since they are BPS. Orthogonal states must remain orthogonal under this change of $g$, so we cannot imagine that all the states become in some sense the `same' black hole; rather there must be $e^{S_{micro}}$ different solutions  which form the ensemble describing the black hole. At large $g$ we can use the gravity description for these states; thus we must find $e^{S_{micro}}$ states of the full quantum gravity theory.

What do these gravity states look like? Let us consider the traditional picture of the black hole. The classical geometry is determined by the mass, charges and  angular momenta so there are no deformations (`hair') that can distinguish the states in the region $r>0$. Thus if the $e^{S_{Bek}}$ states must be orthogonal then they must differ sharply in a planck sized ball around the singularity $r=0$. 

But all such microstates will have a horizon since they all behave like the naive black hole away from $r=0$, and the horizon for each microstate will have approximately the same area $A$ that we would associate to the classical hole. We would then be tempted to associate an entropy $e^{A/4G}$ with the microstate. Does it make sense to have an entropy associated to a microstate? Let us investigate this question in some more detail.

Consider a gas in a box, isolated from its surroundings. The gas is in any {\it one} state out of the possible ensemble of states.
It can still be useful to associate an entropy $S$ to this gas, since we can learn about the behavior of the gas under changes of macroscopic parameters by using  thermodynamic quantities like the free energy $F=E-TS$. In this case the entropy $S$ represents the log of the number of possible microstates which {\it resemble} the state under consideration and therefore are part of the ensemble.

But if we really wanted to study the gas in the box in full detail, we could give the positions and momenta of each of the atoms in the box $\{\vec x_i, \vec p_i\}$. Then we would {\it not} associate an entropy to the microstate. The important point is that we have this freedom -- we can choose to study the full state (with no entropy) or study approximate properties of the state (in which case we have the entropy of the ensemble).

Let us now return to the black hole. For the conventional view of the black hole we have found that each microstate must have a horizon. If we accept the idea that an entropy must be associated to a horizon then we {\it must} associate an entropy to the microstate. This looks strange, since we do not get the option of choosing whether we wish to consider the state as part of an ensemble or not. By contrast, in our picture of microstates we found {\it no horizon} for individual microstates. For this to happen the geometry must change all through the region interior to the horizon, and we found such to be the case. Thus we can describe the microstate fully and do not need to associate any entropy with the microstate. It is true though that the generic microstate is a complicated `fuzzball', so we might prefer to obtain some approximate `averaged' description of the 
interior of the fuzzball. In this case  all the states of the system would together constitute the ensemble with entropy $S_{micro}$. If we draw a boundary of the fuzzball then the area of this boundary satisfies $A/4G\sim S_{micro}$, so we can call it the `horizon'. 

We thus see that the fuzzball picture of the black hole interior is in line with the way we obtain entropy in other physical systems, while the naive black hole geometry is not. It appears sensible that microstates should have no horizons. While we have obtained this picture of the interior for 2-charge states we would expect that a similar picture holds for all holes.

\section{Other results on the `fuzzball' conjecture}

We have given above the simplest results that support the idea that black holes are `fuzzballs'. For lack of space we did not consider other results supporting the conjecture, but we list them briefly here.

\medskip

(i)  {\it Travel times}

\medskip

In the D1-D5 picture we saw that the 2-charge brane bound state could take $e^{2\sqrt{2}\pi\sqrt{n_1n_5}}$ configurations, because the effective string with winding number $n_1n_5$ could be split into `component strings' with winding number $k_i$, $\sum_i k_i=n_1n_5$. (The component strings could also be in different polarization states arising from fermion zero modes.) Consider the special subset (\ref{teight}) where the component strings all have the same winding $k$ and the same polarizations as well. The geometries dual to these CFT states can be constructed; they turn out have a  $U(1)\times U(1)$ axial symmetry. 

Suppose a graviton is incident on such a D1-D5 bound state. In the CFT picture, the graviton can be absorbed into the bound state, creating (in the leading order process) one left and one right mover on one of the component strings. In the dual gravity picture the graviton simply enters the throat of the geometry and starts moving towards $r=0$. It is found that the probabilities for the absorption into the CFT state $P_{CFT}$ and the probability to enter the throat of the supergravity solution $P_{sugra}$ agree \cite{dm}
\be
P_{CFT}=P_{sugra}
\ee
This result in fact led to an understanding of Hawking radiation from near extremal black holes as just a  process of emission from excited branes. But taking the physics a step further we find a potential contradiction. In the CFT deccription the left and right moving excitations travel around the component string (at the speed of light) and re-collide after a time
\be
\Delta t_{CFT}={2\pi R k\over 2}=\pi R k
\label{ttone}
\ee
where $2\pi R k$ is the total length of the component string and the division by $2$ is because each particle need move only half the distance along the string. Thus after a time $\Delta t_{CFT}$ the graviton can be re-emitted from the bound state.
In the gravity description if we had the naive D1-D5 geometry (\ref{d1d5naive}) then the graviton falling down the throat
would keep going down the throat, and not return at all. (This is the case if we do not take backreacktion into account, but it can be shown that even if we include backreaction the quantum will not return for times much longer than (\ref{ttone}) \cite{lm4}.) But the {\it actual} geometries that we find have `caps' at some point down the throat. The wave equation for the graviton can be separated and solved in these axisymmetric geometries, and it is found that the graviton falls down the throat, bounces off the cap, and returns to the start of the throat (where it can escape into the asymptotically flat region) in a time \cite{hot,lm4}
\be
\Delta t_{sugra}=\pi R k=\Delta t_{CFT}
\label{tttwo}
\ee
Thus we observe that if we had the naive geometry then we would not get agreement with the CFT, while with the actual geometries we do get agreement. Note that this is a very detailed agreement, since it works for each value of $k$. The geometries for small $k$ have shallow throats and those for large $k$ have deeper throats, so that (\ref{tttwo}) holds in each case. 

\medskip

(ii) {\it 3-charge states}

\medskip

It would be very interesting if we could analyze the 3-charge D1-D5-P system the same way we analyzed the D1-D5 system. The 3-charge hole is expected to reflect all the physically important properties of black holes, so that lessons learnt for this system will extend to black holes in general. The CFT states for this system are constructed by taking the `component string' picture for the 2-charge D1-D5 states and adding left moving momentum P along the component strings. 

We cannot yet make the geometries dual to the generic 3-charge CFT state. But we have constructed those for some special subsets. The first case is where all component strings have winding $k=1$ and  the same fermion zero modes,  and we have one unit of momentum on one of the component strings. In the dual supergravity this should correspond to a solution of the linearized perturbation equations around the 2-charge D1-D5 background; this perturbation will have to carry $E=P=1/R'$ where $R'$ is the radius of the $S^1$.  It is nontrivial that there should be such a normalizable perturbation: If a solution of the supergravity equations is chosen to be bounded in the `cap' then it will in general diverge at infinity and a solution chosen to be normalizable at infinity would in general be singular somewhere in the interior region. But using a matching technique, carried out to four orders in the the parameter $({\rm length ~of ~throat})^{-1}$, we we find that there does exist a normalizable solution with the right energy and momentum to represent the CFT state \cite{mss}. Since the perturbation is smooth we have found that  this solution carrying D1, D5 and one unit of P charge is smoothly capped, with no horizon or singularity.

The other case we addressed is where we excite {\it each} component string in the same way, by filling up the $n$ lowest allowed fermionic energy levels ($n=1,2,\dots$). In this case we can find the exact geometries dual to the CFT states \cite{gms1,gms2,lunin}. The geometry is again found to end in a smooth `cap' ; there is no horizon or singularity. Thus the geometry is like  Fig.2(b) rather than the naive geometry of a Reissner-Nordstrom extremal hole.
Since all the component strings are chosen to be in the same state, their angular momenta add up and the resulting solution has a significant amount of rotation. These states may thus appear very nongeneric, but we  believe that generic states will have a similar `capped' structure. This is because a very similar behavior happens for the 2-charge case, where we do know all the geometries. If we choose all component strings to have the same winding $k$ and the same fermion zero modes  (eq. (\ref{teight})) then the spins from these zero modes add up, and we get a large angular momentum for the system. The dual geometry turns out to be  generated in the NS1-P frame by the profile (\ref{yyonePrime}) with $\omega=k/(n_1 R)$. Thus the NS1 string carries the momentum by swinging in a uniform helix of $k$ turns. Each unit of momentum thus also contributes a certain amount of angular momentum, and as noted in the CFT picture, the angular momenta all add up to give a significant amount of rotation. These geometries are simpler than the typical geometry  since they are axisymmetric. In the generic geometry the NS1 string will vibrate in different directions at different points, leading to little or no net angular momentum. But the vibrations will still lead to a nonzero size for the system, and thus a `capped' geometry rather than the naive one. Similarly for the 3-charge case we expect that the axisymmetric geometries will be the ones that are easier to find, and while the others are harder to construct they will have the same qualitative features. It turns out that in the 3-charge case the area of tan $r=const$ surface asymptotes to a constant down the throat. This implies that  we are guaranteed to find $A/4G=S_{micro}$ from the area of the fuzzball boundary, since in the naive geometry the area of the horizon at $r=0$  reproduced $S_{micro}$ in the computation of \cite{sv}.

In \cite{mathur} a very rough estimate was made for the 3-charge extremal bound state, and the size of this bound state was found to be of order the horizon radius. In \cite{gm2} an indirect argument was given to support the idea  of `caps' for generic 3-charge geometries. It was shown that  that the size of the fuzzball state was consistent with a relation like (\ref{tttwo}) between the supergravity travel time and the travel time in the 3-charge CFT state. A formalism has been developed \cite{gmr} that allows us to write all classical supersymmetric solutions for the 3-charge case, and the metrics of \cite{gms1,gms2} emerge as particular  cases provided we allow a slight extension of this formalism \cite{gm1}. But we need to identify which of these states are bound states, and it would also be useful to have a map between the geometries and states of the dual CFT.

\medskip

(iii) {\it Supertubes}

\medskip

Consider the NS1-P bound state in IIA string theory, with the NS1,P charges along the direction $y$ as in the constructions above. We can regard this as a configuration in M theory, where a circle $x_{11}$ has been compactified to reach the IIA description. Now we do the $y\leftrightarrow x_{11}$ flip, so that we define IIA string theory through compactification on the circle $y$. The NS1 brane is still  an NS1 brane, but the P charge is now D0 charge. Bound states of the fundamental string and D0 charge were shown to expand to {\it supertubes} \cite{supertubes}. Thus supertubes give another language to study the nontrivial size of 2-charge bound states. The supertube language was then extended to the 3-charge case, where families of 3-charge supertubes were constructed and analyzed \cite{super2}. It is found that these 3-charge configurations can  expand to nontrivial size, and in some cases that were studied this size was shown to be consistent with a `horizon sized' ball.
Other results of interest in related directions are \cite{super3}.

\section{Nature of the fuzzball}

 We have counted the states of the NS1-P system, and found that there are $e^{2\sqrt{2}\pi\sqrt{n_1n_p}}$ states (eq. (\ref{three}). On the other hand we have constructed in the classical limit a continuous family of geometries (\ref{ttsix}), which were parametrized by the vibration profile $\vec F(v)$. In what sense does a quantum state of the NS1-P system correspond to a geometry from this family?

We can ask a similar  question even for an ordinary  string in the laboratory, with no gravity relevant to the physics. The vibrations of the string can be decomposed into harmonics, and the amplitude of each harmonic is quantized like a harmonic oscillator coordinate. To get the energy eigenstates of the system we must excite each oscillator to one of its eigenlevels; this gives the set of different allowed eigenstates. But the eigenstate of an oscillator are Gaussian-type wavefunctions with
some spread, and the overall state of the string is thus some Gaussian-type wavefunction spread around the central axis of the string. As we increase the energy of vibration on the string, we increase the size of the transverse region over which the wavefunction is spread.  Thus the eigenstates are  very `quantum'; in these eigenstates the string does not have a well-defined classical vibration profile.

But we can reach the more familiar `classical profile' of the string in the following way. Suppose that we have a large numb momentum on the string, and that this momentum is partitioned among only a few harmonics. Then the excitation level of the corresponding harmonic oscillators for  is $m_i\gg 1$. For each oscillator we superpose neighboring energy eigenstates 
with appropriate phases to get a coherent state, and this state is a narrowly peaked Gaussian wavefunction  around a steadily oscillating classical mean. The entire string  wavefunction is then a narrowly peaked Gaussian around a classical vibration profile $\vec F(v)$. Note that  if  we had only a few excitations in each harmonic then the corresponding oscillator is in one of its lowest eigenstates, and coherent states do not afford a good classical description. 

We have a similar situation with our NS1-P system. We have a large $n_p$, so we do have a high value of the total vibration level on the string. If we put this energy into just a few of the harmonics then we can use coherent states to describe the dynamics and the classical profile $\vec F(v)$ will be a good description of the string state. In this case the corresponding geometry (\ref{ttsix}) will be a good description of the system; the entire gravity wavefunctional will be narrowly peaked around the classical geometry.  Conversely, we can imagine starting with the continuous family (\ref{ttsix}) and performing a semiclassical quantization of the moduli space generated by the different choices of $\vec F(v)$. 
This will yield a discrete set of quantum states of the gravity theory, which we can relate to the eigenstates of the NS1-P system. Such a quantization has not yet been carried out for the full system with gravity, but in \cite{mp} the moduli space of supertubes (in flat space) was quantized, and the correct count of states was obtained. 

But note that for a {\it generic} state of the NS1-P system the excitation level for each oscillator is $m_i\sim 1$ (see eq. (\ref{ttseven})). This happens because the entropy gets maximized if we split the energy among many different modes, rather than keeping it in a few modes. Thus the generic state will be quite `quantum' and will not be well described by  classical geometry. In other words, there will be sizable fluctuations ${\delta g\over g}$ of the metric, due to the fact that wavefunction is not sharply peaked around a given classical profile. This is why we use the term `fuzzball' to describe the generic state. 

What is of importance to us however is the {\it size} of this fuzzball. This size depends on the mean value of $k$, the wavenumber of the harmonics of the NS1. If $k$ is small, we need a large amplitude to carry the same momentum P, and the transverse size will be big. If $k$ is large we get the same P with a small amplitude of vibration, and the transverse size is small. (In the D1-D5 picture this relates directly to the travel times in (\ref{tttwo}). Small $k$ gives large fuzzball sizes and thus shorter throats, and we get small travel times $\Delta t_{sugra}$ to the cap; large $k$ gives deeper throats and a longer travel time.) The maximal entropy is contributed by excitations with $k\sim\sqrt{n_1n_p}$, and for such $k$ we get the fuzzball radius  (\ref{tfour}). While we can have bigger and smaller fuzzballs (for smaller and larger $k$ respectively) we expect a fairly sharp peak around $k\sim\sqrt{n_1n_p}$ since for large numbers $n_1,n_p$ the entropy  peaks sharply around its optimal parameters, as in any large statistical system. 

To summarize, we can use classical geometries (\ref{ttsix}) to estimate the size of the bound state for a given mean value of $k$. These classical geometries describe well only those states where we put a large number of quanta into a few harmonics, so that the average occupation of each harmonic is $m_i\gg1$. Generic states have $m_i\sim 1$, and so a large quantum fluctuations ${\delta g\over g}$ in the supergravity fields. But the size of the `fuzzball' does not depend on the amount of fluctuation; it just depends on the mean value of $k$. The entropy peaks for $k\sim \sqrt{n_1n_p}$ and this gives the size (\ref{tfour}).

\section{Summary and conjectures}

We have argued, using some examples, that bound states in string theory expand to a size that depends on their degeneracy, so that the area of the `fuzzball' boundary satisfies a Bekenstein type relation $A/4G\sim S$. If such is indeed the case then the black hole interior is nontrivial, and the horizon is not a region of `empty space'. For non-BPS states we can then  imagine that information in the state can be carried out in the radiation from the fuzzball boundary.

What do we see if we fall into such a hole? We cannot answer this question at this stage since we have studied only equilibrium states, not dynamics. But we should note that there are two time scales associated to a black hole: The crossing time across its diameter and the much longer Hawking evaporation time. It is possible that interactions with the `fuzz' are weak enough that an infalling observer does not notice the fuzz as he falls in on the crossing timescale. Over the Hawking evaporation timescale  the black hole plus observer system can relax to its maximal entropic state;  in this process the information in the observer will be slowly integrated into the fuzz, and  his information can thus leak out in the slow radiation process. 

What is the physical reason that states in string theory want to swell up to a size which depends on their degeneracy? One possibility is that this is just an issue of phase space. Consider the NS1-P system, and demand that the state have a very small transverse spread, much smaller than the typical fuzzball radius (\ref{tfour}). What states are allowed? We have seen that if we wish to carry the momentum $n_p$ with a very small transverse amplitude of vibration then we must restrict ourselves to using very high frequencies. We get very few states by partitioning the momentum in this restricted way. 
We can thus see that to accommodate the generic state of the system we must allow the string  a certain transverse volume 
for its fluctuation. This is of course the case for any quantum system: 
We can have only one state of the system per cell of volume $\hbar^d$ in phase space ($2d$ is the dimension of the phase space). The dynamics will typically set some relation between the spread in position and the spread in momentum, so that we cannot get the phase space volume we want by making the position spread arbitrarily small and the momentum spread correspondingly large. For our NS1-P system we might therefore be able to trace the nontrivial size (\ref{tfour}) to the phase space needed to account for the entropy $\sim \sqrt{n_1n_p}$.

If such is indeed the reason for the nontrivial size of fuzzballs then it suggests that black holes give the optimal way to pack a large number of states in a given volume of phase space, and that in any such a packing we cannot reduce the system size
below the radius which we conventionally call the horizon. It would be very interesting to understand why this utilization of phase space gives a universal Bekenstein type  relation between the boundary area and  the entropy. 

It is interesting to see how the 2-charge and 3-charge geometries that we have constructed avoid the no-hair theorems for black holes. {\it If} we have a horizon, then analysis of the gravity field equations at the horizon gives a no-hair theorem. But in the actual geometries the horizon never forms; instead the geometry ends in a `cap'. The naive geometry is spherically symmetric, but all the actual geometries break spherical symmetry. We can take one of these geometries, make a family of geometries that are the same except that they are rotated by some angle, and superpose these geometries to produce a spherically symmetric solution. But since we had to superpose many classical geometries to get this spherical solution, we will have large fluctuations of ${\delta g\over g}$ in the resulting wavefunctional. 

What about the  naive geometry which was classical and spherically symmetric? If it was a smooth geometry, or a geometry that could be shown to be generated by a valid source in the theory, then we would have to accept it as a legitimate solution. The naive solution is singular in all cases. The {\it actual} solutions are completely smooth in the D1-D5 language and are generated by a legitimate source in the NS1-P language. In the latter language we can see why no classical geometry can be spherically symmetric: there is a transverse vibration profile $\vec F(v)$ and this must break the symmetry around $r=0$. We thus find that the spherically symmetric classical naive solution is not a valid solution of the 2-charge system. 

Let us summarize the `fuzzball' picture of black holes suggested by the above results. We have argued that bound states in string theory are not in general planck sized or string sized, but have a size that grows with the degeneracy of the bound state. Perhaps this size is simply a consequence of   the  phase space volume  required to give that degeneracy. In the simple cases we studied we found  that the  bound state is `horizon sized' which means that if we draw a boundary around the typical state then the area of this boundary gives the degeneracy by a Bekenstein type relation. We constructed all states for 2-charges,  and found that the microstates do not individually have a horizon or singularity. The absence of a horizon for individual microstates made sense, since we should be able to view such a microstate without associating any entropy to it. We can call the fuzzball boundary the `horizon'.  This identification corresponds to a kind of `coarse-graining' and so it is consistent that entropy be associated to the area of such a horizon. The generic fuzzball state is rather quantum, but the size of such a state can be estimated by looking at special subsets of states that do have good classical descriptions. We should not try to `average' in some way the states inside the fuzzball region to get a unique classical geometry for the interior; there are $e^S$ orthogonal states of the Hilbert space and they must remain orthogonal in every description. The response of the state to a dynamical process is another matter: It is possible that on the short `crossing timescale' seen by infalling observers the generic fuzzball  states all look the same. What is important is that over the longer Hawking evaporation timescale the differences between states should allow  information to escape in the radiation, and the horizon sized nature of the fuzzball makes this possible.  

\section*{Acknowledgements}

  I would like to thank Stefano Giusto, Oleg Lunin, Ashish Saxena and Yogesh Srivastava who have been collaborators in various aspects of the work discussed here. I also thank Stefano Giusto and Yogesh Srivastava for help with this manuscript. I would like to thank the organizers and participants of the summer school in Crete where this
  talk was given. This work is
supported in part by DOE grant DE-FG02-91ER-40690.

\end{document}